\documentclass[aps,epsfig,psfig,prl,twocolumn,superscriptaddress,showpacs,showkeys,email]{revtex4}
\usepackage{graphics}
\usepackage{graphicx}
\begin{document}
\title{\bf Identifying critical residues in protein folding: Insights from $\phi$-value and $P_{fold}$ analysis}
\author{P.~F.~N. Fa\'\i sca}
\affiliation{Centro de F\'\i sica Te\'orica e Computacional, Universidade 
de Lisboa, Av. Prof. Gama Pinto 2, 1649-003 Lisboa, Portugal}
\email{patnev@cii.fc.ul.pt}
\author{R.~D.~M. Travasso}
\affiliation {Centro de F\'\i sica Computacional, Departamento de F\'\i sica, Universidade de Coimbra, 3004-516 Coimbra, Portugal}
\author{R.~C. Ball}
\affiliation{Department of Physics, University of Warwick, Coventry CV4 7AL, U.~K.} 
\author{E.~I. Shakhnovich}
\affiliation{Department of Chemistry and Chemical Biology, Harvard University, 12 Oxford Street, Cambridge, MA 02138, U.~S.~A.} 
\email{eugene@belok.harvard.edu}

\pacs{\bf{87.15.Cc; 91.45.Ty}}
\keywords{\bf{mutagenesis, native geometry, kinetics, folding pathways}}

\begin{abstract}

We apply a simulational proxy of the $\phi$-value analysis and perform extensive
mutagenesis experiments to identify the nucleating residues in the folding `reactions' of two small lattice G\=o polymers with different native geometries. Our findings show that
for the more complex native fold (i.e., the one that is rich in non-local, long-range bonds), 
mutation of the residues that form the folding nucleus lead to a considerably larger increase
in the folding time than the corresponding mutations in the geometry that is predominantly local. These results are compared with data obtained from an accurate analysis based on the reaction coordinate folding probability $P_{fold}$, and on structural clustering methods. Our study reveals a complex picture of the transition state ensemble. For both protein models, the transition state ensemble is rather heterogeneous and splits-up into structurally different populations. For the more complex geometry the identified subpopulations are actually structurally disjoint. For the less complex native geometry we found a broad transition state with microscopic heterogeneity. These findings suggest that the existence of multiple transition state structures may be linked to the geometric complexity of the native fold. For both geometries, the identification of the folding nucleus via the $P_{fold}$ analysis agrees with the identification of the folding nucleus carried out with the $\phi$-value analysis. For the most complex geometry, however, the apllied methodologies give more consistent results than for the more local geometry. The study of the transition state' structure reveals that the nucleus residues are not necessarily fully native in the transition state. Indeed, it is only for the more complex geometry that two of the five critical residues show a considerably high probability of having all its native bonds formed in the transition state. Therefore, one concludes that in general the $\phi$-value correlates with the acceleration/deceleration of folding induced by mutation, rather than with the degree of nativeness of the transition state, and that the `traditional' interpretation of $\phi$-values may provide a more realistic picture of the structure of the transition state only for more complex native geometries.  
\end{abstract}
\maketitle

\section{Introduction}

The folding kinetics of the vast majority of small, single domain proteins is remarkably well
modeled by a two-state process, where the unfolded state (U) and the native fold (N) are separated by a high free energy barrier, on the top of which lays the transition state (TS)~\cite{JACKSON}. Due to its transient nature, the structural characterization of the folding TS represents a particularly challenging task in protein biophysics. Indeed, experimental studies to date have typically relied on the application of a particular class of protein engineering methods, the so-called $\phi$-value analysis, pioneered by Fersht and co-workers in the 1980s~\cite{FERSHT0}. In the $\phi$-value analysis a mutation is made at some position in the protein sequence; 
the $\phi$-value is obtained by measuring the mutation's effect on the folding rate and stability, namely $\phi = -RT ln(k_{mut}/k_{WT})/\Delta \Delta G^{N-U}$, where $k_{mut}$ and $k_{WT}$ are the folding rates of the mutant and wild-type (WT) sequences respectively, and $\Delta \Delta G^{N-U}$ is the free energy of folding.  For a non-disruptive mutation (i.e., a mutation that does not change the structure of the native state, and does not alter the folding pathway either), $-RT ln(k_{mut}/k_{WT})$ can be approximated by the change in the activation energy of folding upon mutation, $\Delta \Delta G ^{TS-U}$, and therefore 
$\phi = \Delta \Delta G ^{TS-U}/ \Delta\Delta G ^{N-U}$.\par
A $\phi$ value of unity means that the energy of the TS is perturbed on mutation by the same amount the native state is perturbed, which has traditionally been taken as evidence that the protein structure is folded at the site of mutation in the TS as much as it is in the native state. Conversely, residues which are unfolded in the TS, as much as they are in the unfolded state, should exhibit $\phi$ values of zero. The interpretation of a fractional $\phi$ value is, however, not straightforward as it might indicate the existence of multiple folding pathways~\cite{FERSHT1, DILL}, or it may underlie a unique TS with genuinely weakened interactions~\cite{FERSHT1}.
An alternative interpretation of mutational data has been recently proposed by Weikl and co-workers that instead of considering the effect of each individual mutation, collectively considers all mutations within a fold's substructure (e.g., a helix). Such an interpretation is able to capture the so-called nonclassical $\phi$ values ($\phi<$ 0 or $\phi> 1$), and explains how different mutations at a given site can lead to different $\phi$ values~\cite{MERLO, WEIKL}.\par 
In the case of the 64-residue protein chymotrypsin-inhibitor 2 (CI2), the extensive use of $\phi$-value analysis revealed only one residue (Ala 16) with a distinctively high $\phi \sim 1$, whereas the vast majority of CI2's residues show typically low fractional $\phi$-values~\cite{FERSHT2}. 
These findings were taken as evidence that CI2 folds via the so-called nucleation-condensation (NC) mechanism, with the folding nucleus (FN) consisting primarily of the set of bonds (mostly local but also a few long-range) established by the residue with the highest $\phi$-value~\cite{FERSHT1, FERSHT3}, which is identified as a nucleation site. Interestingly, the very first microscopic evidence for the existence of a nucleation mechanism in protein folding was obtained in the scope of Monte Carlo simulations of a simple lattice model, where Shakhnovich and collaborators observed that once the FN, consisting of a specific set of native bonds, is established the native fold is achieved very rapidly~\cite{ABKEVICH}. Additional studies {\it in vitro} \cite{MILLA, LOPEZ, KIEFHABER, FULTON} and {\it in silico}~\cite{DAGGETT, SHAKHNOVICH_2001, VENDRUSCOLO, FERNANDEZ_2002, SHAKHNOVICH_2002, AKANUMA_2005, SHEA_2005, PACI_2006, SHAKHNOVICH_PNAS_2006, TRAVASSO_2007a, TRAVASSO_2007b}, using more sophisticated protein models and other simulational methodologies, have provided further evidence for the existence of nucleation sites in CI2 as well as in other target proteins. For this reason the nucleation mechanism is typically considered the most common folding mechanism amongst small, two-state proteins~\cite{NOLTING}.\par  
A few years ago, S\'anchez and Kiefhaber reported a set of experimental data indicative 
that $\phi$-values are considerably inaccurate unless the difference in the folding free energy upon mutation is larger than 7kJ/mol~\cite{SANCHEZ}. A refute by Fersht followed 
based on the premise that the 7kJ/mol cut-off was based on mutations that are unsuitable for $\phi$-value analysis because they are disruptive~\cite{FERSHT4}. More recently, a collaborative effort between three laboratories in North America investigated the relationship between $\phi$-value reliability and the change in the free energy of folding, $\Delta\Delta G ^{N-D}$, using the generally employed experimental practices and conditions. A conclusion came out from this study stating that the precision of experimentally determined $\phi$-values is poor unless $\Delta\Delta G ^{N-D}> 5$ Kcal/mol~\cite{RIOS}. In a related study, Raleigh and Plaxco pointed out that only three out of the 125 more accurately determined $\phi$-values reported in the literature lie above 0.8, and that about 85\% of the mutations characterized for single domain proteins show $\phi$-values below 0.6~\cite{RALEIGH}. Overall, these findings have prompted a discussion regarding the existence of specific nucleation sites, and therefore some controversy has been generated regarding the nucleation mechanism of protein folding.\par 
The goal of the present study is to contribute to clarify this controversy by applying two different procedures that identify the nucleating residues in the folding of small lattice proteins. One of these procedures, a simulational proxy of the $\phi$-value analysis, leads to a supposedly `inaccurate' identification of the FN's residues that is made irrespective of the free energy changes caused by mutation. The other procedure, which is based on the use of the reaction coordinate, $P_{fold}$~\cite{DU}, allows for an accurate/rigorous identification of the TS, and of the native contacts which make up the FN. By comparing the results obtained from both approaches insight is gained on the suitability of the $\phi$-value analysis as a tool to identify kinetically determinant residues in protein folding, and on the nucleation mechanism of folding.\par
Since native geometry is known to play a major role in the folding kinetics of small two-state proteins~\cite{PLAXCO, GROMIHA, ZHOU, FAISCA_2002, FAISCA_2004, FAISCA_2005}, we study two model proteins with considerably different native geometries.\par
This article is organized as follows. In the next section we describe the protein models and computational methodologies used in the simulations. Afterwards, we present and discuss the results. In the last section we draw some concluding remarks.      

\section{Models and Methods}

\subsection{The G\={o} model and simulation details}

We consider a simple three-dimensional lattice model of a protein molecule with
chain length $N$=48. In such a minimalist model amino acids, represented by beads of uniform size, occupy the lattice vertices and the peptide bond, which covalently connects amino acids along the polypeptide chain, is represented by sticks with uniform (unit) length corresponding to the lattice spacing. \par 
To mimic protein energetics we use the G\={o} model~\cite{GO}. In the G\={o} model the energy of a conformation, defined by the set of bead coordinates $\lbrace \vec{r_{i}} \rbrace$, is given by the contact Hamiltonian  
\begin{equation}
H(\lbrace \vec{r_{i}} \rbrace)=\sum_{i>j}^N
\epsilon \Delta(\vec{r_{i}}-\vec{r_{j}}),
\label{eq:no1}
\end{equation}
where the contact function $\Delta (\vec{r_{i}}-\vec{r_{j}})$, is unity only 
if beads $i$ and $j$ form a non-covalent native contact (i.e., a contact  
between a pair of  beads that is present in the native structure) and is zero otherwise. 
The G\={o} potential is based on the principle that the native fold is very 
well optimized energetically. Accordingly, it ascribes equal stabilizing 
energies (e.g., $\epsilon=-1.0 $) to all the native contacts and neutral energies 
($\epsilon =0$) to all non-native contacts. 
In order to mimic the protein's relaxation towards the native state we use a 
Metropolis Monte Carlo (MC) algorithm~\cite{METROPOLIS, CHAN_DILL, MAREK_HOANG} together with the kink-jump move set~\cite{BINDER}. A MC simulation starts from a randomly generated unfolded conformation 
and the folding dynamics is monitored by following the evolution of the fraction of native
contacts, $Q=q/L$, where $L$ is number of contacts in the native fold and $q$ is the 
number of native contacts formed at each MC step. The number of MC steps required to fold to the native state (i.e., to achieve $Q=1.0$) is the first passage time (FPT) and the folding time, $t$, is computed as the mean FPT of 100 simulations. Except otherwise stated folding is studied at the so-called optimal folding temperature, the temperature that minimizes the folding time~\cite{OLIVEBERG2, JCPSHAKH, CIEPLAK, PFN1}. The folding transition temperature, T$_{f}$, is defined is the temperature at which denatured states and the native state are equally populated at equilibrium. In the context of a lattice model it can be defined as the temperature at which the average value $<Q>$ of the fraction of native contacts is equal to 0.5~\cite{ABKEVICH1}. In order to determine T$_f$ we averaged $Q$, after collapse to the native state, over MC simulations lasting $\sim$ 10$^9$ MCS. \par

\subsection{Target geometries}

\begin{figure}
{\rotatebox{0}{\resizebox{7cm}{7cm}{\includegraphics{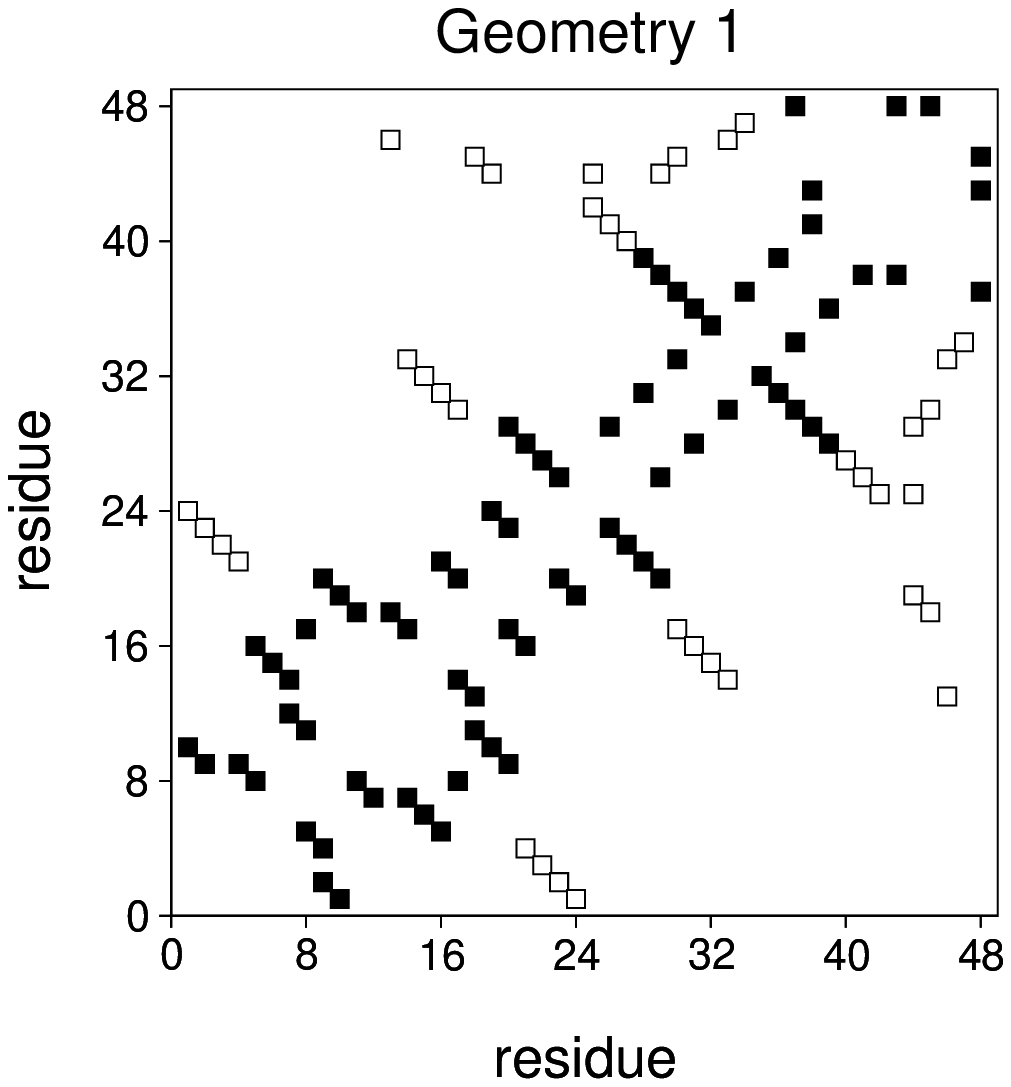}}}}\\
{\rotatebox{0}{\resizebox{7cm}{7cm}{\includegraphics{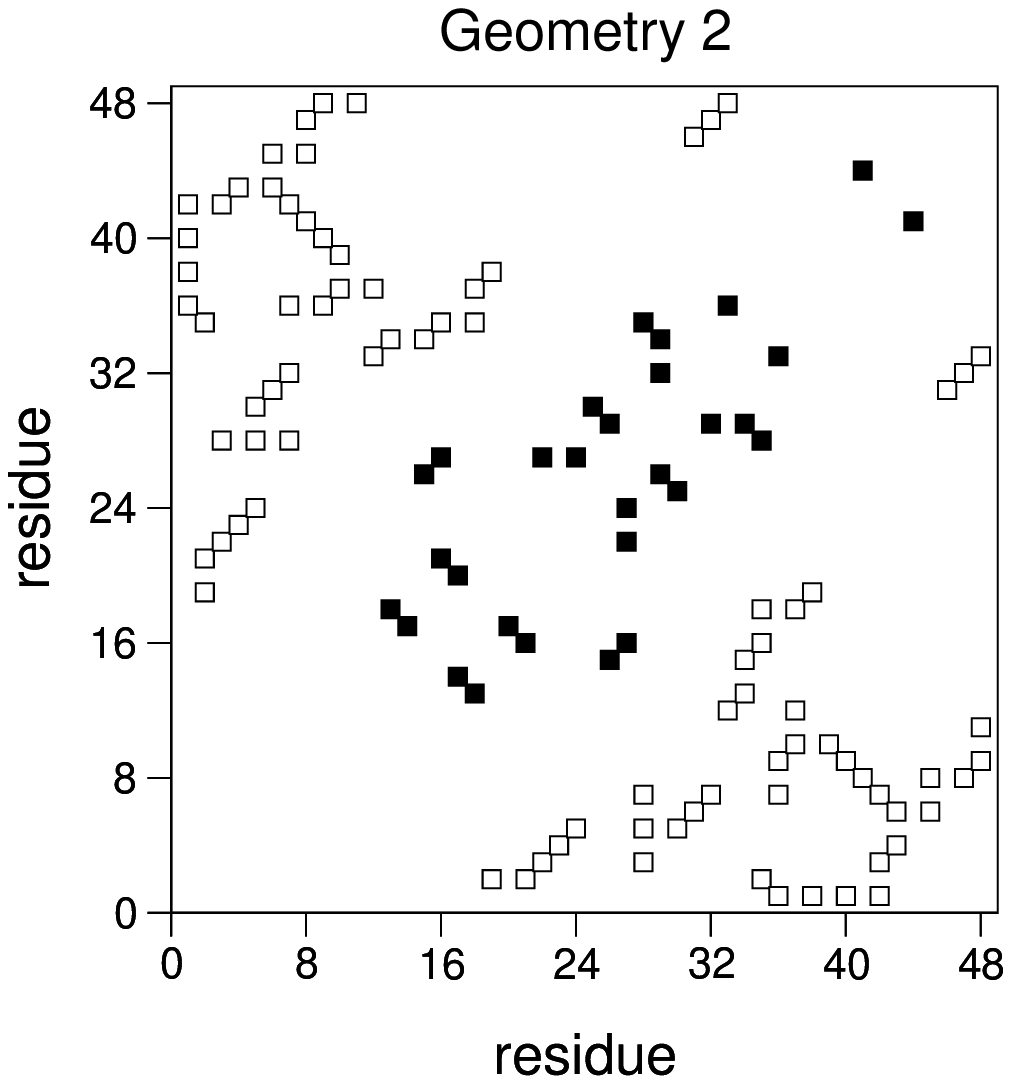}}}}
\caption{Geometries 1 and 2 represented through their contact maps. Each square represents a native contact. For structures that like ours are maximally compact cuboids with $N=48$ residues there are 57 native contacts. A non-local contact between two residues $i$ and $j$ is defined as LR if their sequence separation is at least 12 units, i.e. $|i-j| \geq 12$~\cite{GROMIHA}. Accordingly, the number of LR (white squares) contacts in geometry 1 is 19 and in geometry 2 is 42. The predominance of LR contacts in geometry 2 leads to considerably higher values of the long-range order~\cite{GROMIHA} (0.88 vs. 0.40 for geometry 1) and (absolute) contact order~\cite{PLAXCO} parameters (21.4 vs. 9.9 for geometry 1).}
\label{fig:no1}
\end{figure}

Two native folds, which are amongst the `simplest' (geometry 1) and the most `complex' (geometry 2) cuboid geometries found through lattice simulations of homopolymer relaxation, were considered in this study. A contact map representation, which emphasizes their distinct geometrical traits, is shown in Figure~\ref{fig:no1}.  Table~\ref{table1} provides a summary of kinetic and thermodynamic features of both protein models.

\begin{table}
\caption{Summary kinetic and thermodynamic properties of the protein models considered in this study. E is the native state's energy, T is the optimal folding temperature and $log_{10}(t)$ is the logarithmic folding time computed at T. Also shown is the folding transition temperature, T$_{f}$.}
\begin{ruledtabular}
\begin{tabular}{c c c c c }
geometry   &   E    &      T     &       $log_{10}(t)$  & T$_{f}$   \\ \hline \hline
 1         &   $-57.00$  &    $0.66$  & $5.64 \pm 0.04$ & 0.762  \\
 2         &   $-57.00$  &    $0.67$  & $6.29 \pm 0.05$ & 0.795  \\
\end{tabular}
\label{table1}
\end{ruledtabular}
\end{table}

\subsection{Folding probability}

The folding probability, $P_{fold}(\Gamma)$, of a conformation
$\Gamma$, is defined as the fraction of MC runs which, starting from 
$\Gamma$, fold before they unfold~\cite{DU}. It was shown in the context of 
lattice models that $P_{fold}$ features the appropriate characteristics 
for a reaction coordinate. Accordingly, conformations that are members of the transition 
state have $P_{fold}=1/2$, while pre- and post-transition state conformations have smaller and larger folding probabilities respectively.\par
\begin{figure}
{\rotatebox{0}{\resizebox{8cm}{8cm}{\includegraphics{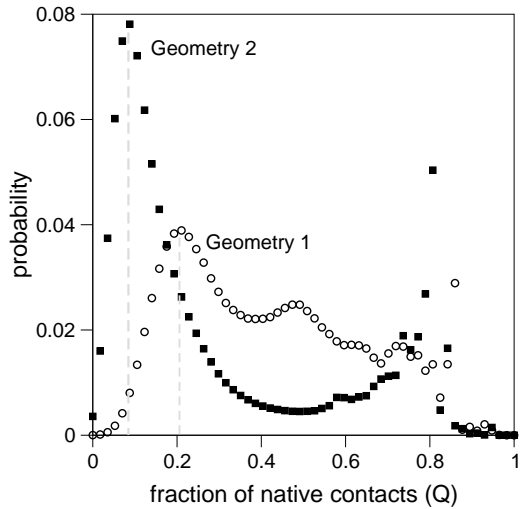}}}}
\caption{Probability distribution for the fraction of native contacts,
$Q$, for geometry 1 and geometry 2  as a function of 
$Q$. A conformation is considered unfolded when $Q < Q_U$ ( $Q_U$ is indicated by the dotted line). } 
\label{fig:no2} 
\end{figure}

Because a $P_{fold}$ calculation amounts to a Bernoulli trial, the relative error resulting from using $M$ runs scales as $M^{-1/2}$~\cite{HUBNER_JMB}. Thus, in order to accurately compute $P_{fold}$ we consider 500 MC runs divided equally into five sets  of 100 folding simulations. The average value of $P_{fold}$ is computed for each set,  and the mean of all five sets, together with its standard deviation, is evaluated. Each MC run stops when either the native fold ($Q=1.0$) or some unfolded conformation is reached. A conformation is deemed unfolded when its fraction of native contacts $Q$ is smaller than some cut-off, $Q_{U}$. In order to estimate $Q_{U}$ we compute the probability of finding some fraction of native contacts $Q$ as a function of $Q$ in 200 MC folding runs (Figure~\ref{fig:no2}). A high-probability peak, centred around the fraction of native contacts $Q=0.2$,  is readily apparent in the graph reported for geometry 1. In the case of geometry 2 the highest probability peak appears around $Q=0.1$. These fractions of native contacts are considerably low and therefore identify states with minimal residual structure. In this work we use these fractions of native bonds to establish the cut-off value $Q_U$ for each model protein.\par
A total of 8000 conformations was collected from 8000 independent MC folding runs, each conformation being sampled from the run's last $5 \times 10^6$ MCS. The folding probability of each conformation was measured as outlined above and conformations were partitioned into seven ensembles with  $P_{fold}=0.2, 0.3,..., 0.8$, each ensemble containing approximately 400 conformations.

\subsection{Structural clustering analysis}

Here we summarize a graph-theoretical method, similar to that described in
Refs.~\cite{HUBNER} and ~\cite{DONALD}, which is used to cluster conformations within each $P_{fold}$ 
ensemble based on their structural similarity. The measure of structural similarity, $r$, between 
two conformations is the number of native bonds they have in common normalized to the maximum 
number of native bonds in the pair. {\it Every} possible pair of conformations is considered in each $P_{fold}$ ensemble. 
Two conformations are structurally similar  ({\em i.e.}, linked) if $r$ is larger than a cut-off $R$, which is fixed so that the largest cluster, the so-called giant component, contains approximately half of the conformations in the starting ensemble. Two
conformations belong to the same cluster if they are linked by a path of connected conformations.

\section{Results}

\section{Identifying critical residues with a simulational proxy of the $\phi$-value analysis}

\subsection{$\phi$-value dynamics}
The mechanistic equivalent of the $\phi$-value of residue $i$ at time $t$, $\phi_{i}^{mec}(t)$, is defined as the ratio between the number of native bonds  $q_{i}^{\Gamma}(t)$ residue $i$ establishes in some conformation $\Gamma$ at time $t$, and the number of bonds $q_{i}^{fold}$ it establishes in the native fold,  $\phi_{i}^{mec}(t)= q_{i}^{\Gamma}(t)/q_{i}^{fold}$~\cite{VENDRUSCOLO, HUBNER_JMB, PACI_07}.\par
Since the formation of the foldin nucleus (FN) is the rate-limiting step in two-state folding, the residues that belong to the FN  will remain in their native environment during a small fraction of the overall folding time. In other words, for a FN's residue $i$, $\phi_{i}^{mec}(t)$ is likely to attain the value $1$  only very close to folding into the native state, and 
for most values of $t$ $\phi_{i}^{mec}(t)$ will be smaller than one.  
Moreover, as a result of structural correlations driven by chain connectivity, residues that are covalently bonded to FN's residues in the polypeptide chain should behave
in a similar way. A similar behavior is also expected for the two terminal residues and their respective neighbours in the chain.\par

\begin{figure*}
{\rotatebox{0}{\resizebox{8cm}{8cm}{\includegraphics{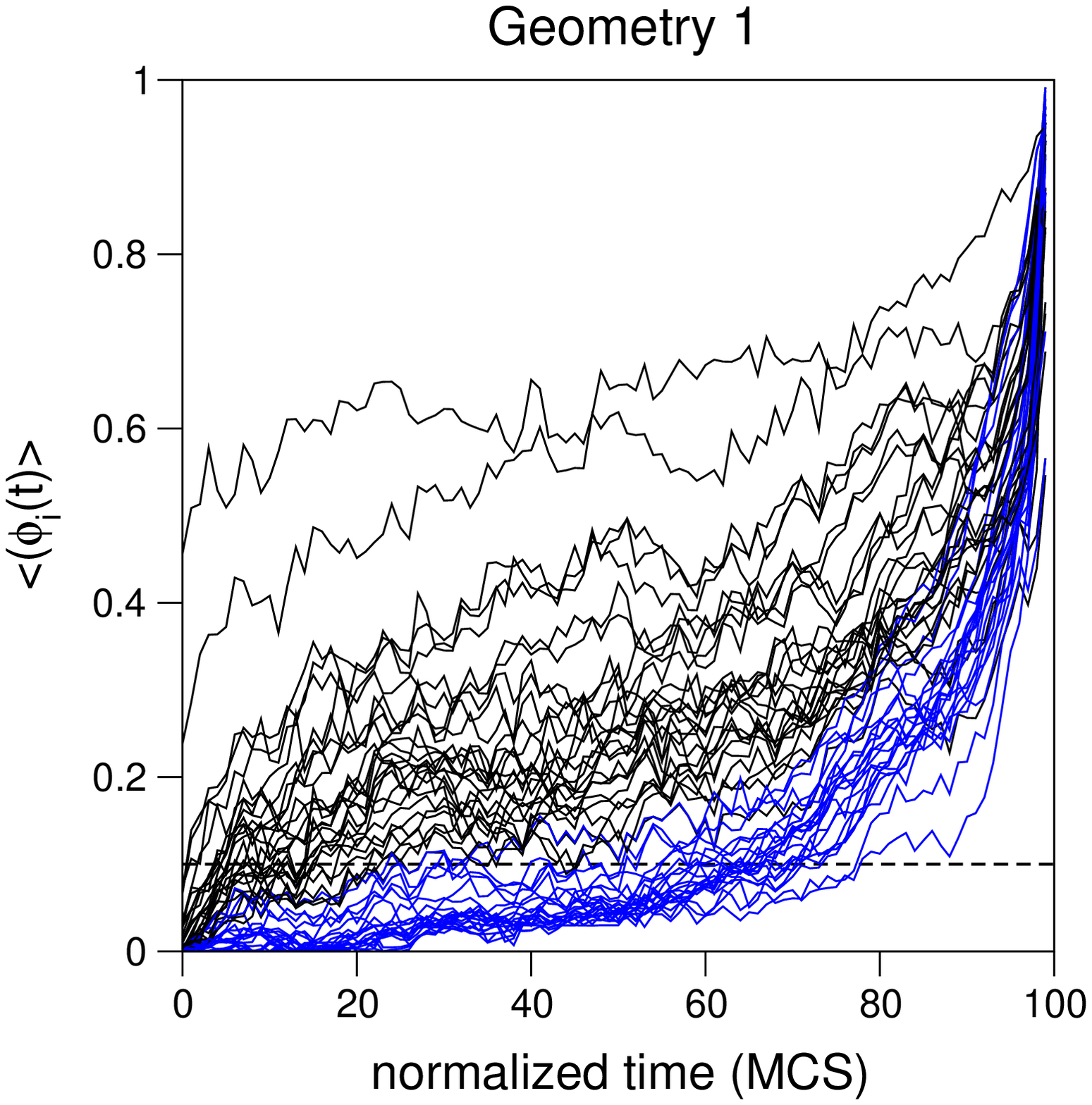}}}}
{\rotatebox{0}{\resizebox{8cm}{8cm}{\includegraphics{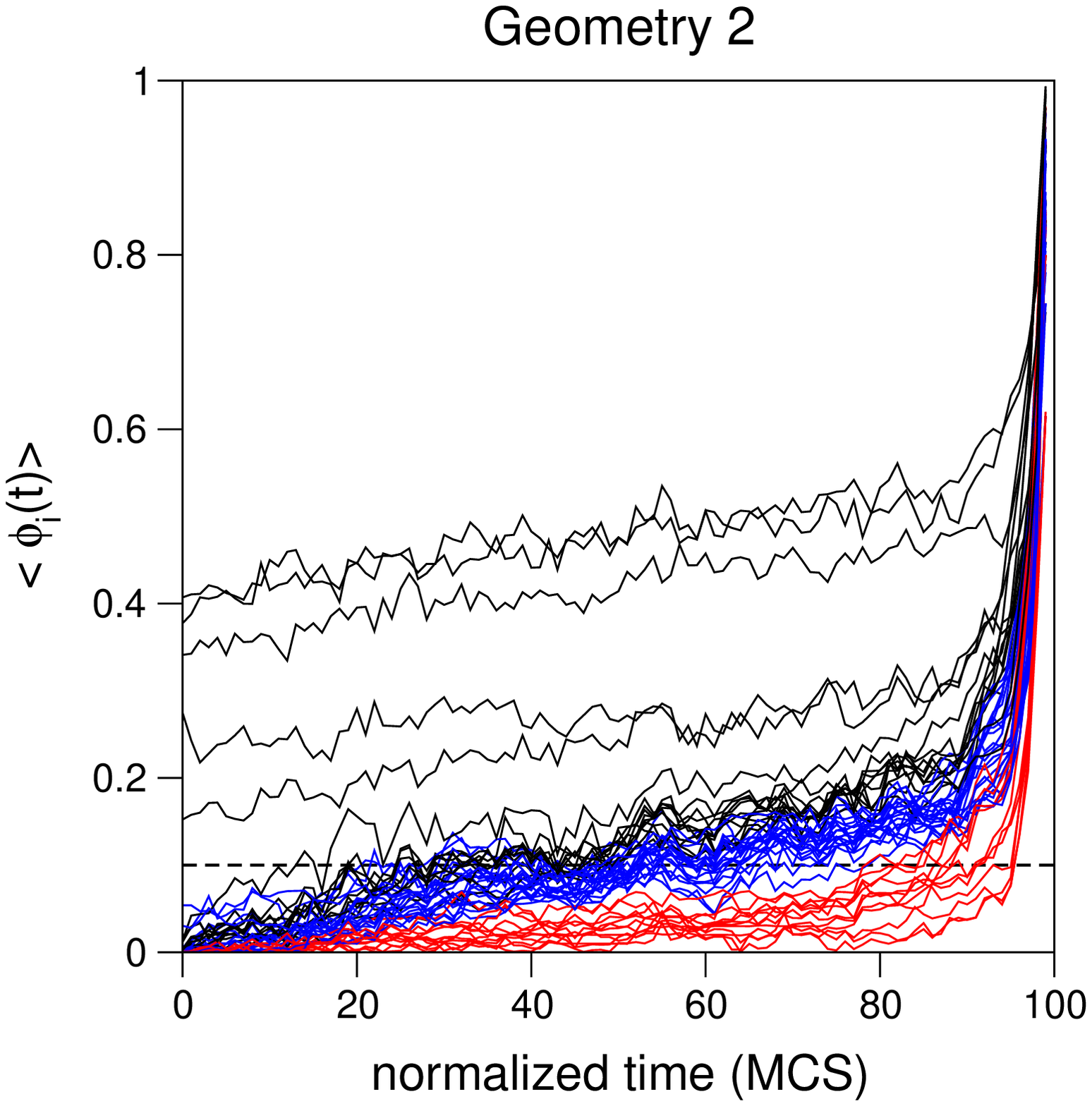}}}}
{\rotatebox{0}{\resizebox{8cm}{8cm}{\includegraphics{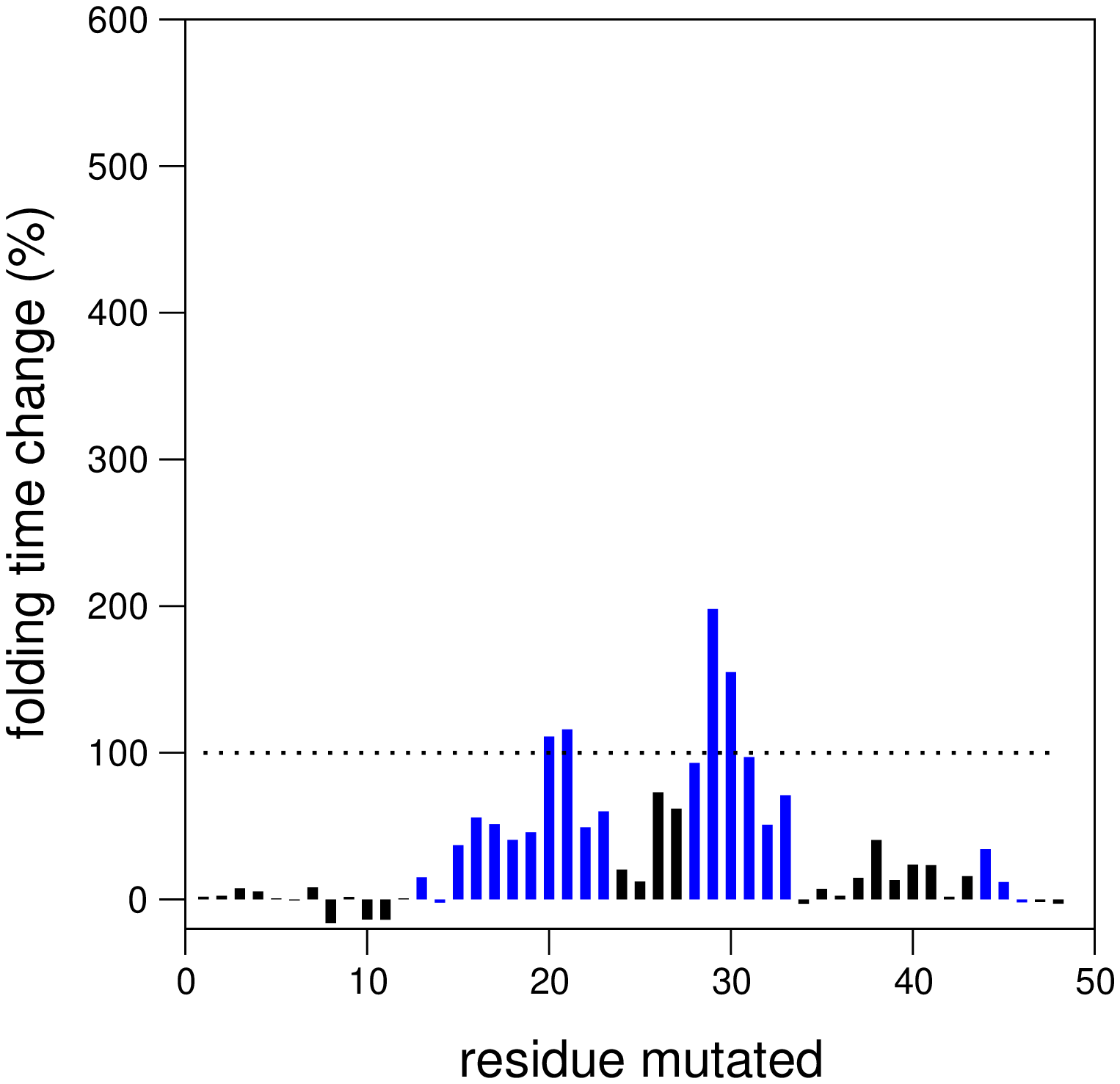}}}} 
{\rotatebox{0}{\resizebox{8cm}{8cm}{\includegraphics{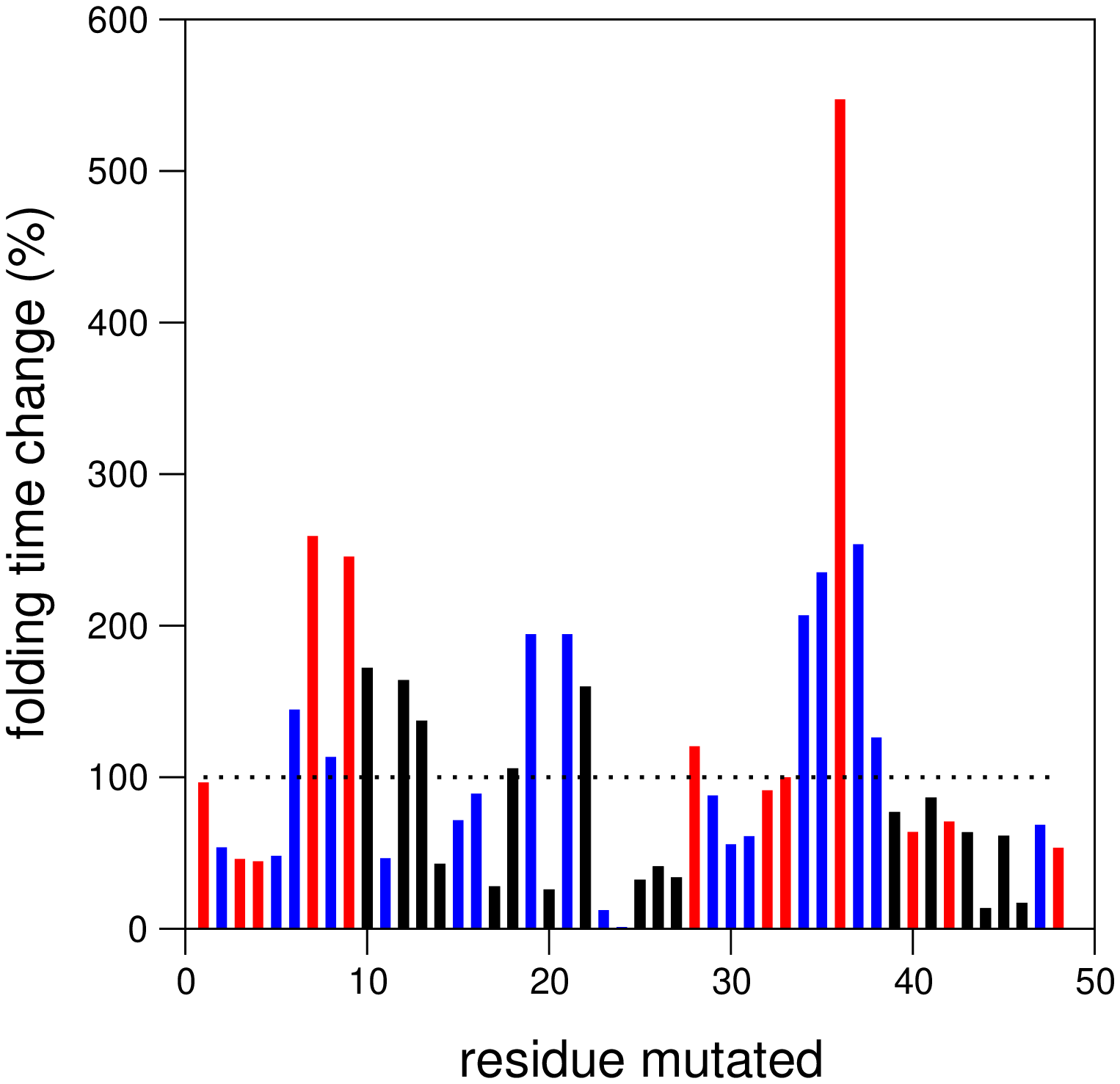}}}}
\caption{Time evolution of the mechanistic equivalent of the $\phi$-value, $\phi_{i}^{mec}(t)$ (top), and change in folding time, relative to the WT sequence, upon performing a single-point mutation.} 
\label{fig:no3}
\end{figure*}

In order to investigate how $\phi_{i}^{mec}(t)$ evolves during folding we proceed as follows. An ensemble of 100 folding runs is considered, and each folding run is divided in 100 bins of length $\Delta t$=FPT$/100 $ MCS. The 100 time  bins correspond to a normalized integer time coordinate $k$ that goes from $0$ to $100$ in all the MC runs.  
For each individual residue and each run, the time average $\phi_{ik}^{mec}$ of $\phi_{i}^{mec}(t)$ when $t$ is in the k-th bin is computed. Then, the averages $\phi_{ik}^{mec}$ for each residue are averaged over the 100 MC runs.  Results obtained for both geometries are reported in Figure~\ref{fig:no3} (top), where the blue curves refer to the residues for which the average value of  $\phi_{i}^{mec}$ is smaller than 0.1, at least during 50\% of the time, and increases  to unity only very late in folding. The red curves in the graph of geometry 2 report residues for which  the average value of  $\phi_{i}^{mec}$ is smaller than 0.1, at least during 90\% of the time, and increases very sharply only late in folding. \par
A qualitatively global analysis of the two sets of 48 curves shows interesting differences. For example, the average value of $\phi_{i}^{mec}(t)$ is considerably much lower  for geometry 2 than for geometry 1. Also, for geometry 2, the curves within each identified subset are closely matched together, which may be taken as an indication that the corresponding residues make and break bonds in a rather independent manner (i.e., bonds form and break more cooperatively in geometry 2 than in geometry 1).

\subsection{Site-directed mutagenesis}

If the formation of the FN is the rate-limiting step in folding, site-directed mutations on the nuclear core residues are supposed to have a significant effect on the folding rate
(or alternatively in the folding time)~\cite{FERNANDEZ}. Therefore a comparison of different mutations is important to identify which particular residues are involved in the TS~\cite{CLEMENTI}. 
In the G\={o} model interactions between residues can either be neutral or stabilizing. Likewise, a single-site mutation within the context of the G\={o} model is equivalent to replacing the set of native bonds established by one residue with neutral bonds (i.e., bonds to which zero energy is ascribed). Because the amino acid sequence is not changed, as in a mutagenesis experiment with real proteins, in principle, one can study the influence of the native contacts without changing the native structure, and without {\it significantly} changing the folding pathway.\par 
Site-directed mutations were performed for every individual residue, and the folding time of the mutant evaluated. The percent change in folding time (relative to the WT sequence) is reported in Figure~\ref{fig:no3} (bottom), where different colours have been used to establish a link with the residue's $\phi$-curves. There is a striking difference between both geometries considered here, which regards the considerably larger folding times observed for geometry 2. Of note, there are several (neutral) mutations (e.g., on residues 1, 2, 6, 12) that do not change the folding time of the WT sequence in geometry 1. Moreover, for geometry 1, there are also a few `abnormal' mutations (e.g., on residues 8, 10 and 11) that actually lead to a decrease of WT protein's folding time. In general, for both geometries, the mutations that lead to a larger increase in the folding time are on the residues that spend a very little amount of time in their native environment during folding.\par
To proceed with the identification of the FN we combine data from both experiments described above and investigate {\it only} the dynamics of the subset of residues that: i. spend (on average) less than 10\%  of time in their native environment during in folding (blue and red bars in Figure~\ref{fig:no3}, bottom) and, ii. whose mutation leads to an increase of at least 100\% in the folding time. A residues satisfying conditions 1 and 2 is deemed potential nucleation site (PNS). 

\subsubsection{Geometry 1}

We have performed double-point mutations by combining {\it all} the pairs of residues which were 
identified as PNSs, and selected only those mutants whose folding
time is larger than that observed for the most deleterious (i.e., severe) single-point mutation. Several double mutants have folding times that are more than one order of magnitude larger than that of the WT sequence. A particularly large increase (of 1.6 orders of magnitude) in the folding time is observed when residues 30 and 20 were simultaneously mutated (Table~\ref{table_mutG1} in the Appendix). 
To proceed with the identification of the FN we have considered triple-point mutations and, perhaps not surprisingly, we have found that residues 20 and 30 participate in three of the most deleterious mutations of this kind, which also involve residue 21 and  residue 29 (Table~\ref{table_mutG1} in the Appendix). Thus, according to the `$\phi$'-value analysis, the nucleating residues for geometry 1 are residues 20, 21, 29 and 30, and the set of bonds they establish is the FN for this geometry.\par

\subsubsection{Geometry 2}

A procedure identical to that used for geometry 1 was applied to geometry 2 that revealed the kinetic relevance of residues 7, 34, 35, 36 and 37. Indeed, the folding times registered upon (double-point) mutating the pairs of residues 35 and 36, 37 and 7, 7 and 34, as well as residues 36 and 7 all lead to folding times that are at least 1.4 orders of magnitude larger than that displayed by the WT protein (Table~\ref{table_mutG2} in the Appendix). Furthermore, several triple-point mutations combining these residues lead to extraordinary high folding times, which are up 2.3 orders of magnitude larger (for residues 7, 34 and 37) than that of the WT sequence, or even folding failure (for residues 7, 35, 36 and residues 7, 35, 37) (Table~\ref{table_mutG2} in the Appendix). These findings are suggestive that for this model protein, the nucleating residues are residues 7, 34, 35, 36 and 37.\par

\section{Identifying critical residues with $P_{fold}$-analysis}

\subsection{Folding pathways}

\begin{figure*}
{\rotatebox{0}{\resizebox{9cm}{4.5cm}{\includegraphics{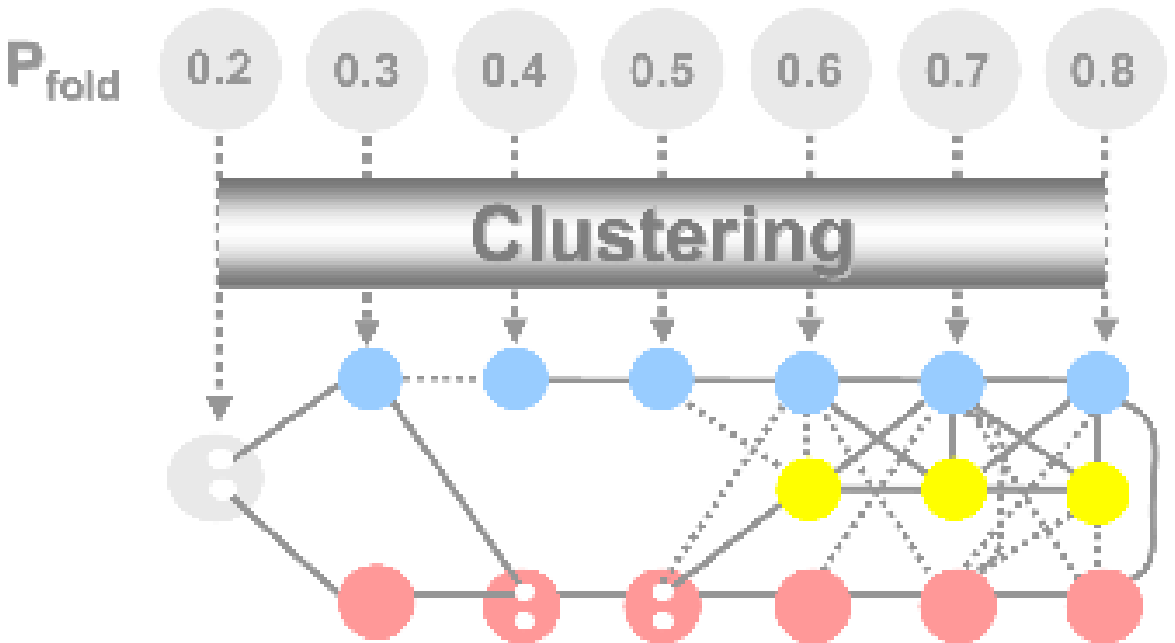}}}}\\
\vspace{1.5cm}
{\rotatebox{0}{\resizebox{9cm}{5.5cm}{\includegraphics{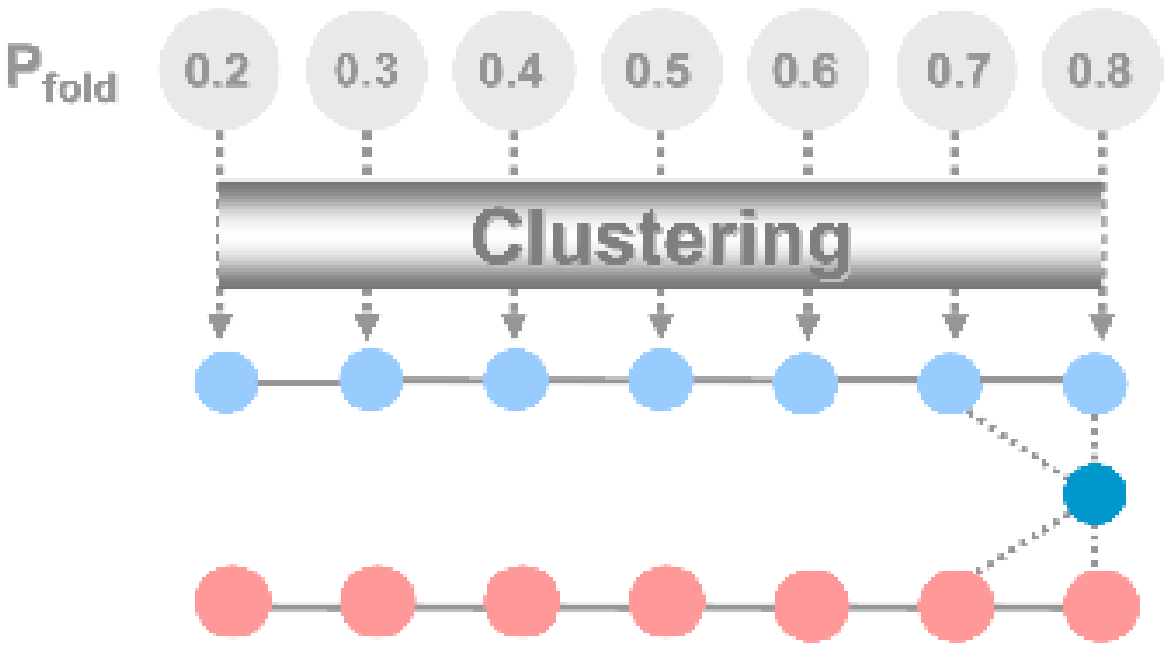}}}}

\caption{Structural classes found along the reaction coordinate $P_{fold}$ for geometry 1 (top) and geometry 2 (bottom). In both graphs the dominant cluster (i.e., the giant component) is show on the bottom row, while the subdominant cluster is represented on the top row. For geometry 2, a trapped state appears at $P_{fold}=0.8$, while for geometry 1 a third cluster, of size similar to that of the subdominant cluster, develops from $P_{fold}=0.6$ onwards. Two clusters are connected through a dotted line if (on average) their conformations share between 0.45 and 0.55 native bonds. A full line is drawn between clusters if their average similarity parameter, $\bar{r}$ is larger than 0.55. Alternatively, no line is drawn between clusters if $\bar{r}<0.45$.}
\label{fig:no4}
\end{figure*}

A folding pathway is a sequence of conformational changes leading to the native structure starting from some unfolded conformation. In order to identify potentially relevant conformational states in the folding `reactions' of geometries 1 and 2 we have applied the structural clustering method previously outlined to several ensembles of conformations with $P_{fold}$ ranging from 0.2 (early folding) to 0.8 (late folding), which were collected from 8000 independent folding 
trajectories. Two clusters of relevant size, named hereafter the dominant cluster (or giant component) and subdominant cluster (this is the largest cluster after the giant component), emerge at successive values of $P_{fold}$ for both geometries (Figure~\ref{fig:no4}). For geometry 1, a third cluster of size similar to that of the subdominant cluster emerges from $P_{fold}=0.6$ onwards (Table \ref{clusters1} in the Appendix). Also, for certain $P_{folds}$ and after the subdominant cluster segregates from the starting ensemble, it is possible to discriminate between two considerably different conformational states within the dominant cluster by applying further clustering to conformations therein. We name these distinct conformational states, dominant cluster 1 and dominant cluster 2. Such a `refining' of the clustering process helps to reveal the more complicated intertwining folding pathways to geometry 1.  
\par
A set of conformations is detected late (at $P_{fold}=0.8$) in the folding to geometry 2 that corresponds to a trapped state. Indeed, folding simulations starting from these conformations last for approximately the same time as folding simulations starting from random-coil type conformers, and are one order of magnitude slower than simulations starting from other conformations having the same $P_{fold}$ (Table \ref{clusters2} in the Appendix). This is possibly a direct consequence of 
the fact that non-native contacts form with a high probability ($>70\%$) in these high-$P_{fold}$ conformers. For this geometry, we have also found that folding simulations  starting from conformations in the subdominant cluster are, for all considered  values of  $P_{fold}$, systematically faster than folding simulations starting in conformations pertaining to the dominant cluster (Table \ref{clusters2} in the Appendix). The difference in folding speed attained is particularly striking in the case of TS and pre-TS conformations with $P_{fold}=0.4$ (Table \ref{clusters2} in the Appendix). This is, however, not surprising  because conformations that belong to those subdominant clusters have the vast majority of their LR native bonds formed with very high probability, which means they have already surmounted most of the entropic cost of establishing LR bonds. On the other hand, conformations in the dominant clusters, while having about the same fraction of native bonds ($Q\approx 0.40$) formed as conformations in the subdominat clusters, still lack the vast majority of their LR contacts whose formation slows down folding.\par

We have determined the mean value of the similarity parameter, $\bar{r}$, between two clusters by averaging the structural similarity parameter $r$ between every pair of conformations (one from each cluster). In Figure~\ref{fig:no4}, two clusters are connected by a full line if $\bar{r} \geq 0.55$, while those for which  $0.45 \leq \bar{r} < 0.55$ are linked with a dotted line. No line is drawn between clusters if $\bar{r} < 0.45$. Along the successive values of $P_{fold}$ the resemblance between clusters of the same type (e.g., between subdominant clusters) is typically larger than the resemblance between clusters of different types. This is particularly evident in the case of geometry 2. For geometry 1, however, once the TS is crossed, a considerable amount of structural similarity develops between clusters of different types. \par
To accurately establish the existence of folding pathways it is necessary to determine if the successive $P_{fold}$ clusters are dynamically linked. A folding pathway exists if a conformation within a cluster can be reached from (at least) one conformation pertaining to a cluster of lower $P_{fold}$. In this case, since the successive dominant (and subdominant) clusters are, on average, very similar to each other (as shown by the high values of $\bar{r}$), it is perhaps straightforward for a conformation in the dominant (subdominant) cluster at $P_{fold}=0.2$ to develop into a conformation in the dominant (subdominant) cluster at $P_{fold}=0.3$, and so on. Therefore, we assume the existence of a set of microscopic folding pathways (to simplify let us name it folding route 1) linking the dominant clusters, and of another set of microscopic folding pathways (folding route 2) linking the subdominant clusters. Folding routes 1 and 2 are parallel if no conformation within a certain $P_{fold}$ cluster in one route can lead to a conformation within any cluster (of larger $P_{fold}$) in the other route. For geometry 2, we have found that starting folding from TS conformations, folding routes 1 and 2 are indeed parallel tracks to the native state. Indeed, 100\% of folding runs starting from conformations in the subdominant cluster at $P_{fold}=0.5$ lead to conformations in the equivalent cluster at $P_{fold}=0.8$ prior to unfolding (i.e., without having to pass through an unfolded conformation).  
Similarly, 90\% of the simulations that start from conformations in the dominant cluster at $P_{fold}=0.5$ end up in conformations within the dominant cluster at $P_{fold}=0.8$, the remaining 10\%  developing into conformers representative of the trapped state (Table \ref{runclusters2} in the Appendix). A completely different scenario holds for geometry 1 where, once the TS is crossed, conformations within folding route 1 evolve into conformers of folding route 2. For example, although 75\% of the conformations in the TS's subdominant cluster develop into conformations in the subdominant cluster at $P_{fold}=0.8$, 11\% of the folding runs end up leading to conformers in the third cluster, and 14\% of the runs end up in the dominant cluster. A similar crossing between pathways is observed for folding runs starting from conformations belonging to the dominant cluster (Table \ref{runclusters1}). Therefore, in the case of geometry 1, the folding routes linking dominant and subdominant clusters are not parallel routes to the native state.\par

\subsection{The structural and geometric characterization of the transition state}

\begin{figure*}
{\rotatebox{270}{\resizebox{7cm}{7cm}{\includegraphics{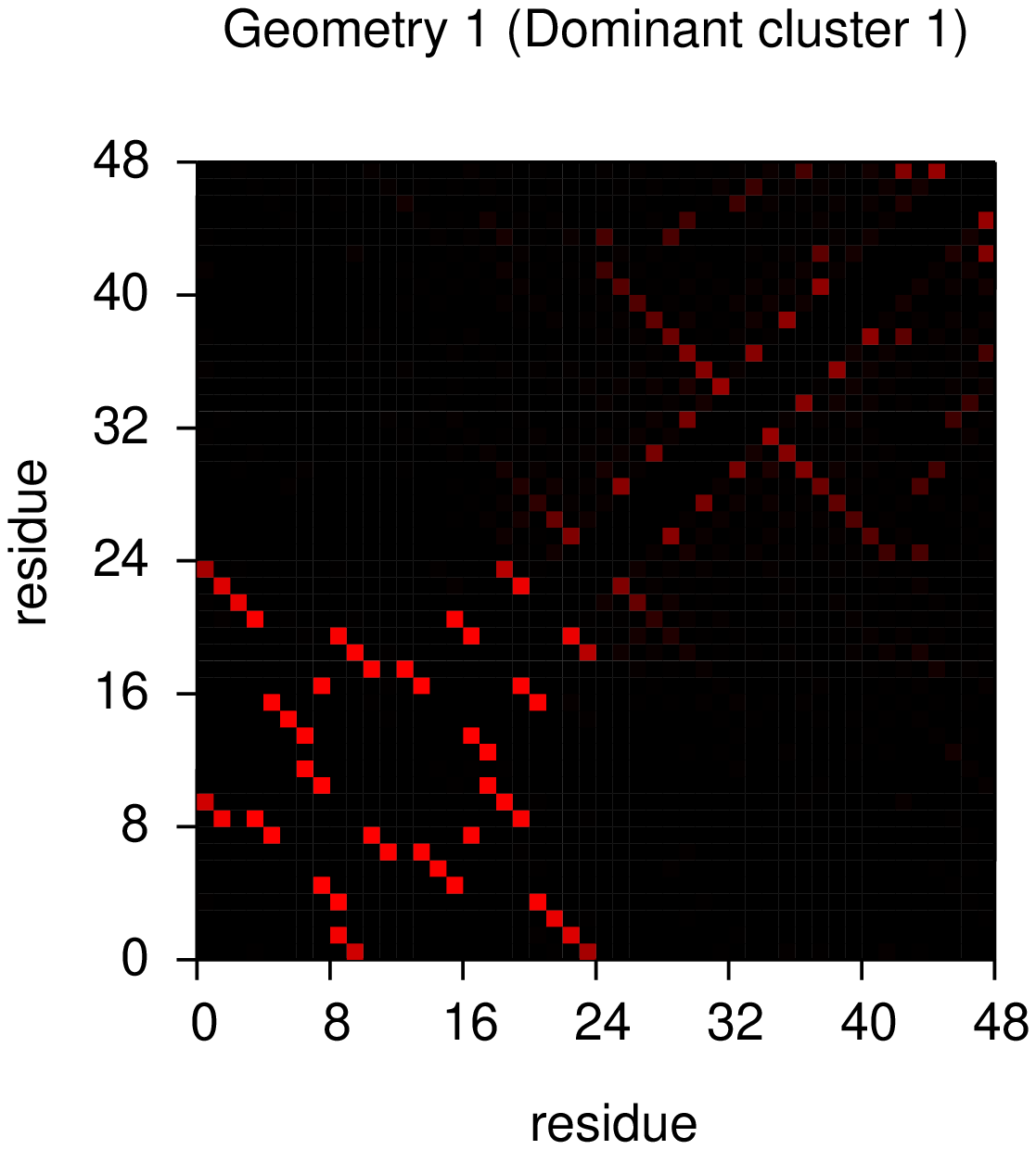}}}} 
{\rotatebox{270}{\resizebox{7cm}{7cm}{\includegraphics{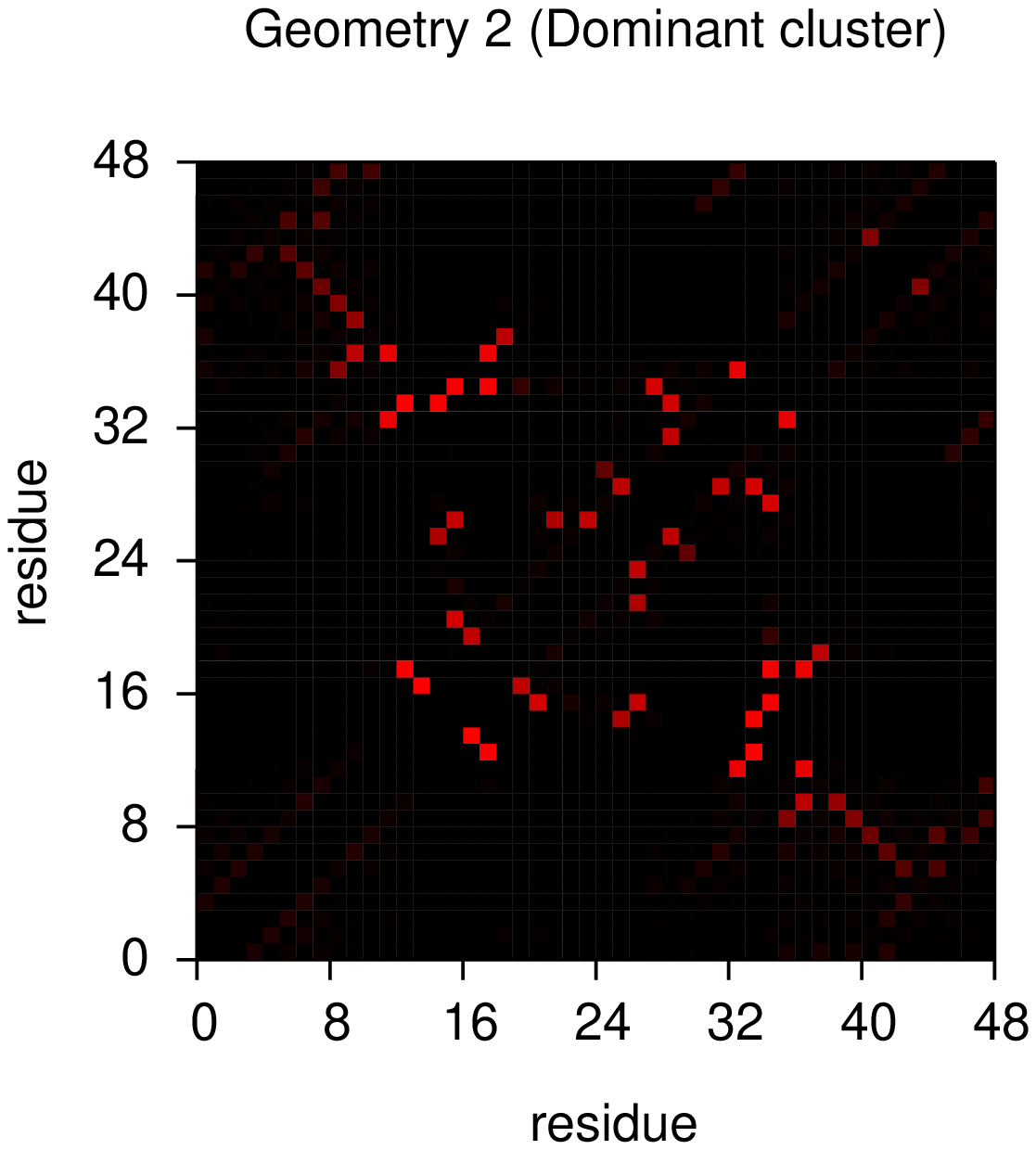}}}} \\
{\rotatebox{270}{\resizebox{7cm}{7cm}{\includegraphics{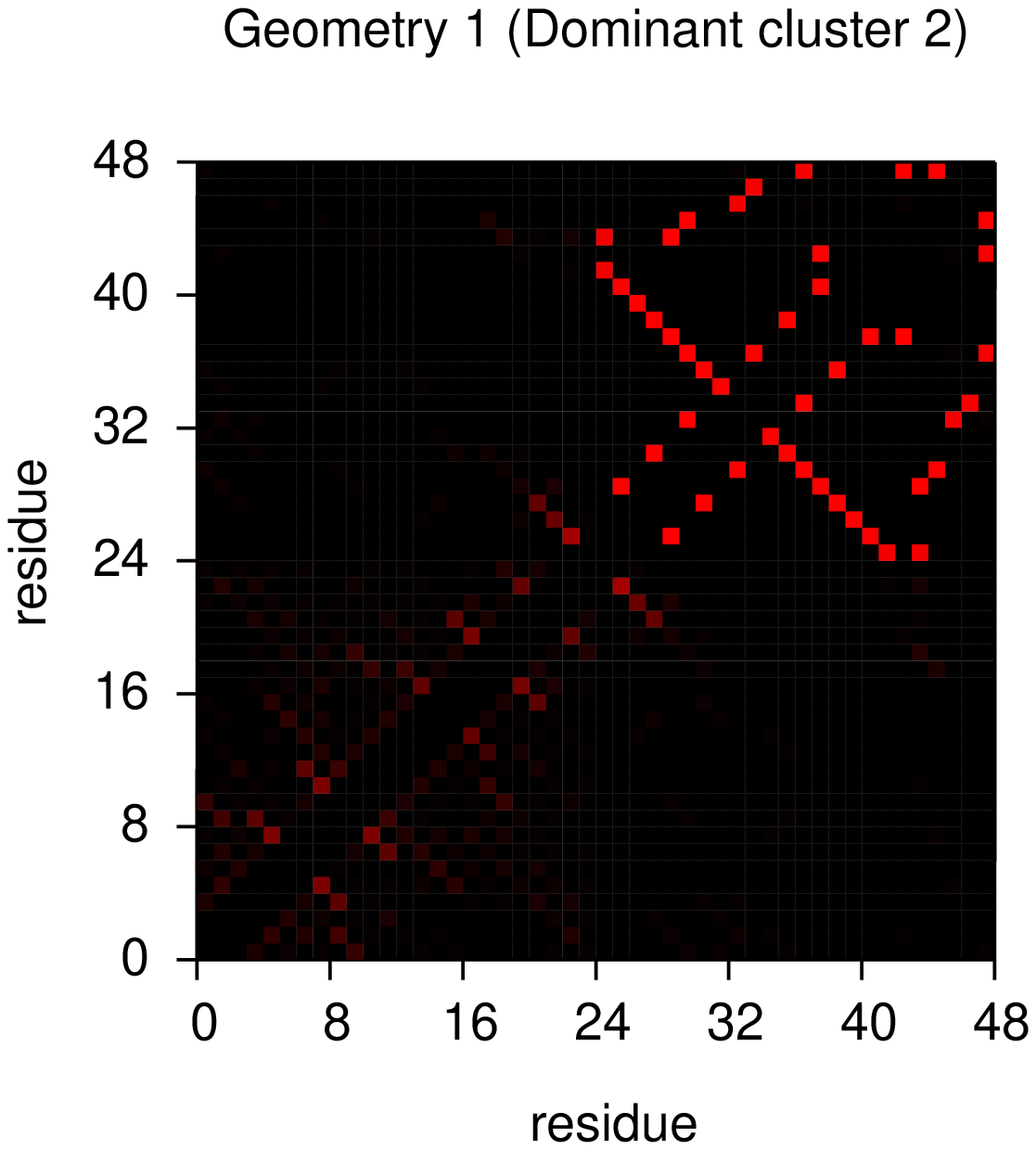}}}} 
{\rotatebox{270}{\resizebox{7cm}{7cm}{\includegraphics{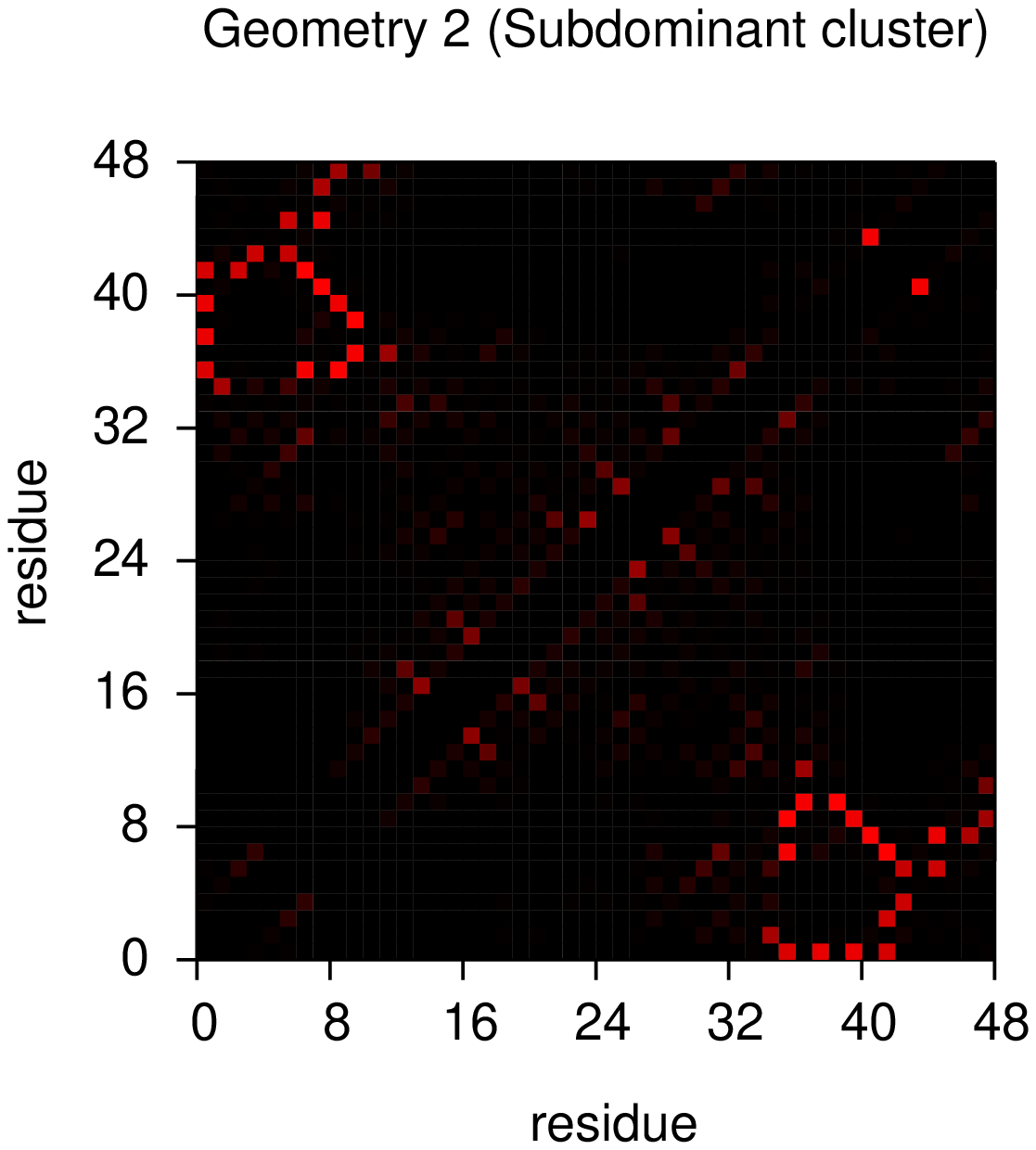}}}} \\
{\rotatebox{270}{\resizebox{7cm}{7cm}{\includegraphics{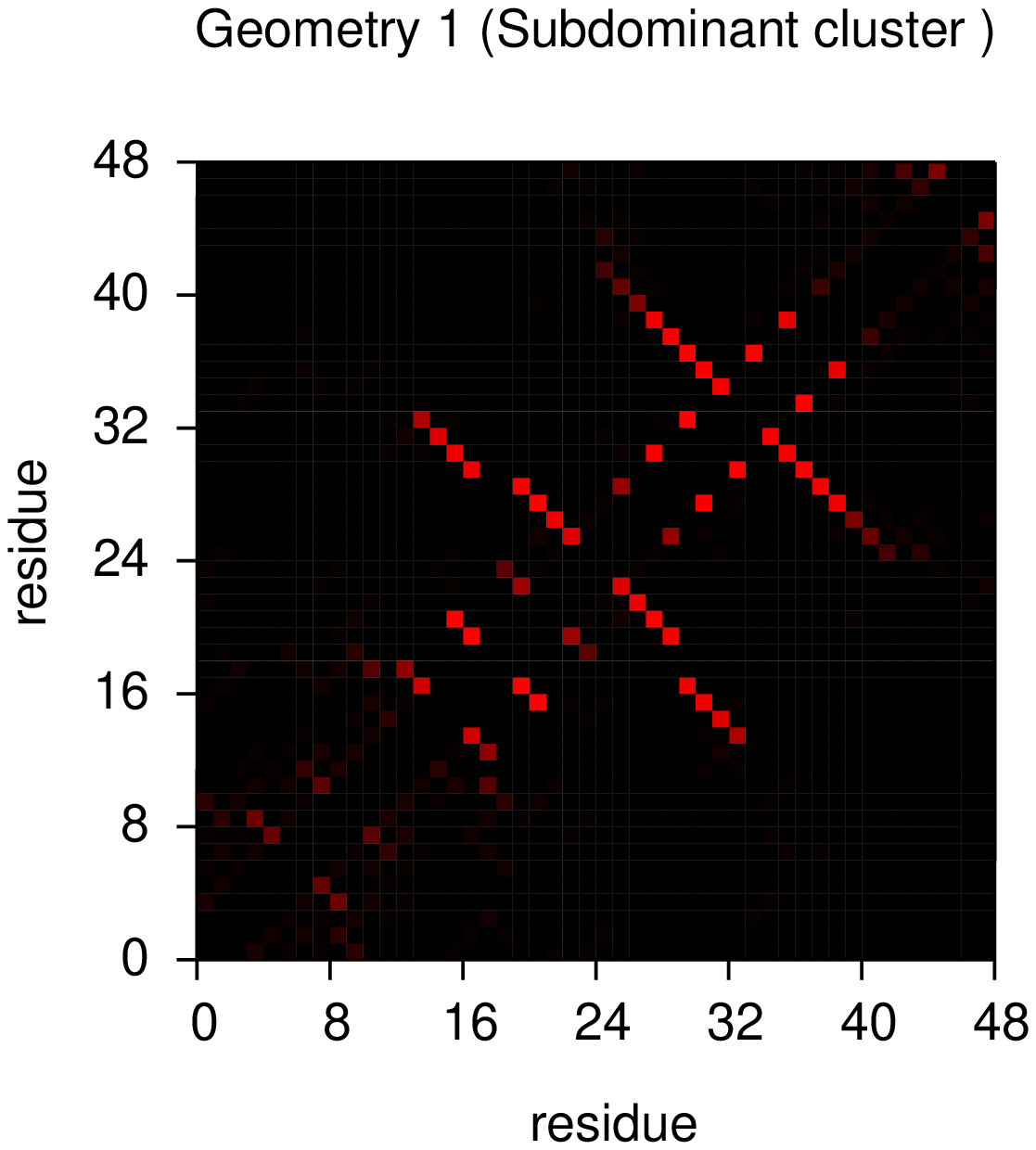}}}} 
{\rotatebox{270}{\resizebox{8cm}{7cm}{\includegraphics{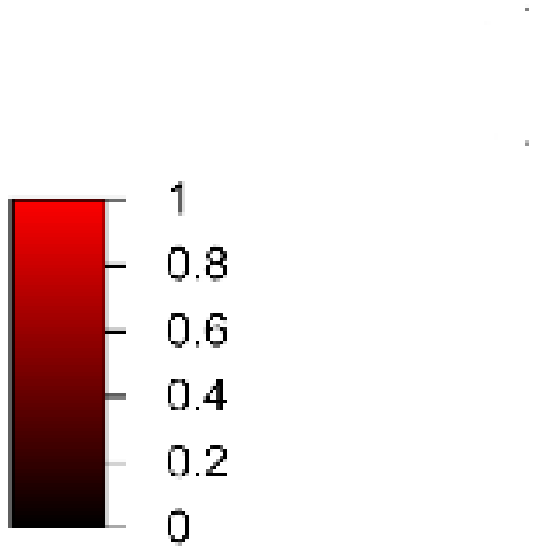}}}} 
\caption{Transition state contact maps for geometry 1 (left) and geometry 2 (right). The probability of forming a native bond in the two structural classes identified within the dominant cluster 1 and 2 of geometry 1 is shown in the left and middle panels respectively. The right panel shows the probability of forming a native bond in the TS's subdominant cluster. For geometry 2, the structure of the TS's dominant cluster is reported on the left panel, while that of the subdominant cluster is shown on panel on the right.}
\label{fig:no5}
\end{figure*}

The structural characterization of the TS comprises not only the identification of the multiplicity of the pathways leading to it but also the degree of structural diversity in 
the ensemble itself. A meaningful discussion of whether the TS is considered to be heterogeneous with alternative forms depends on the resolution at which two such structures differ within the ensemble. Sosnick {\it et al.}~\cite{SOSNICK} propose the following three classes of transition state heterogeneity: 1) a single essential TS nucleus with some partially formed interactions, 2) a structurally heterogeneous ensemble where some residues are critical for the FN but different groups of structures exist at the TS (i.e., conserved FN with microscopic heterogeneity), and 3) the nuclei can be structurally disjoint, each with a diverse set of necessary structures comprising distinct nuclei.
To proceed with the structural characterization of the TS ensemble of the two model proteins considered in the present study we have evaluated the probability that a native contact is formed in both the dominant and subdominant clusters at $P_{fold}=0.5$. 
For the TS ensemble of geometry 2 (Figure~\ref{fig:no5}, bottom) two structural classes can be clearly distinguished. The conformations pertaining to the subdominant cluster
are characterized for having a well-defined group of  $\approx$ 20 non-local LR bonds that form with a considerably high probability ($>0.7$); in these conformers the probability of forming any of the remaining native bonds is, on the contrary, vanishingly small.
Such a `non-local' structural class has an average absolute contact order of ACO=22.7, which is about 106\% of the native structure's ACO. This particular subset of LR bonds has a small probability of forming in the dominant cluster's conformations, which are, for this very reason `local' conformations. Such a larger number of local bonds naturally translates into a smaller average ACO of 15.5, which is about 72\% of the native fold's ACO. Thus, the two TSs structures identified for geometry 2, more than structurally disjoint, are actually structurally  complementary. For geometry 1, there is not such a striking structural difference between dominant and subdominant clusters. Indeed, not only they have a balanced amount of local and non-local bonds (which naturally reflects in their average ACO of 8.3, 8.3 and 7.3, for the dominant cluster 1 and 2, and subdominant cluster respectively), but the subdominant cluster actually shares with the two identified structures of the dominant cluster about 25\% of its highly probable native bonds. This picture is suggestive of the presence of one broad structural class, representative of a single FN, which is structurally heterogeneous.\par

\subsection{The folding nucleus}

\begin{figure*}
{\rotatebox{0}{\resizebox{7cm}{7cm}{\includegraphics{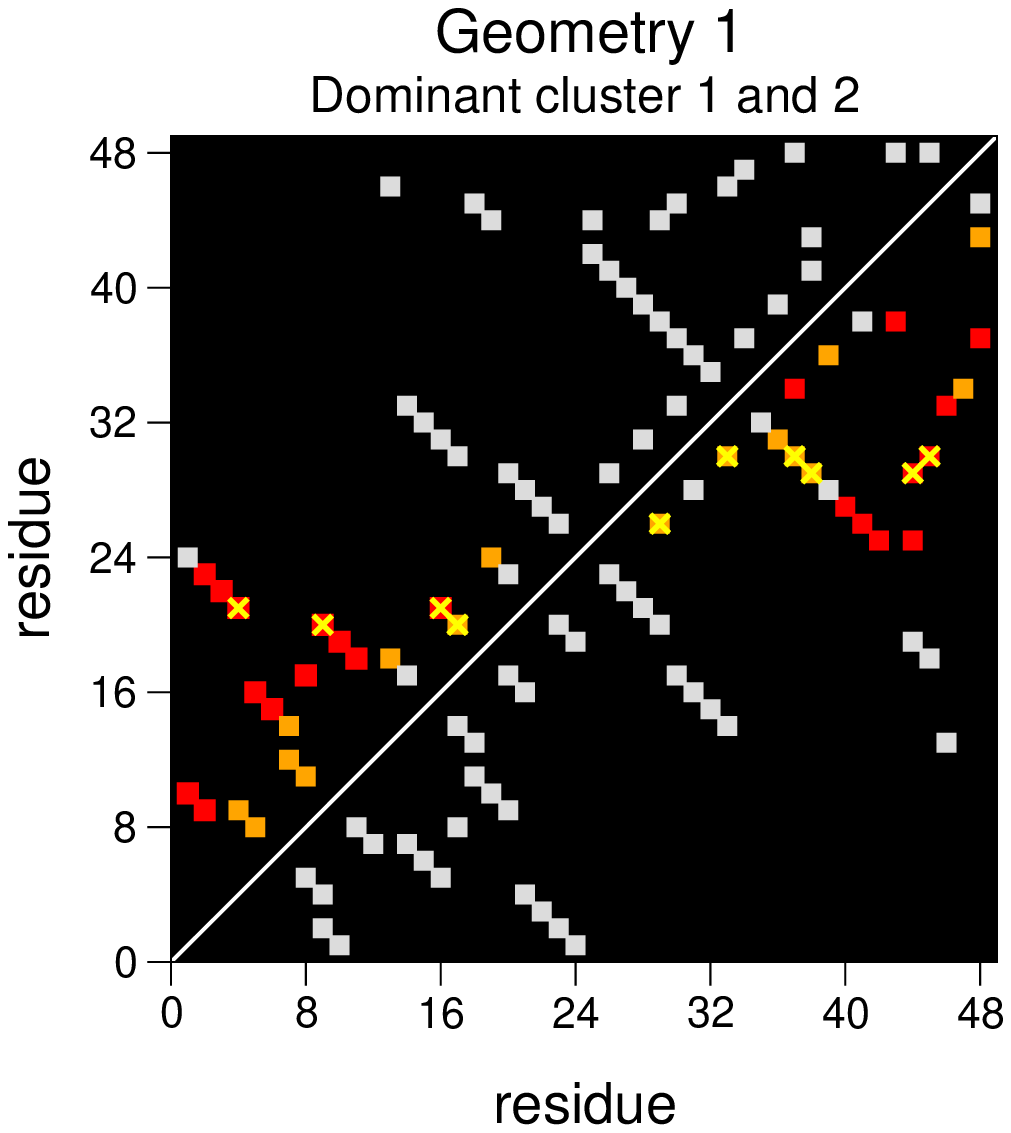}}}}
{\rotatebox{0}{\resizebox{7cm}{7cm}{\includegraphics{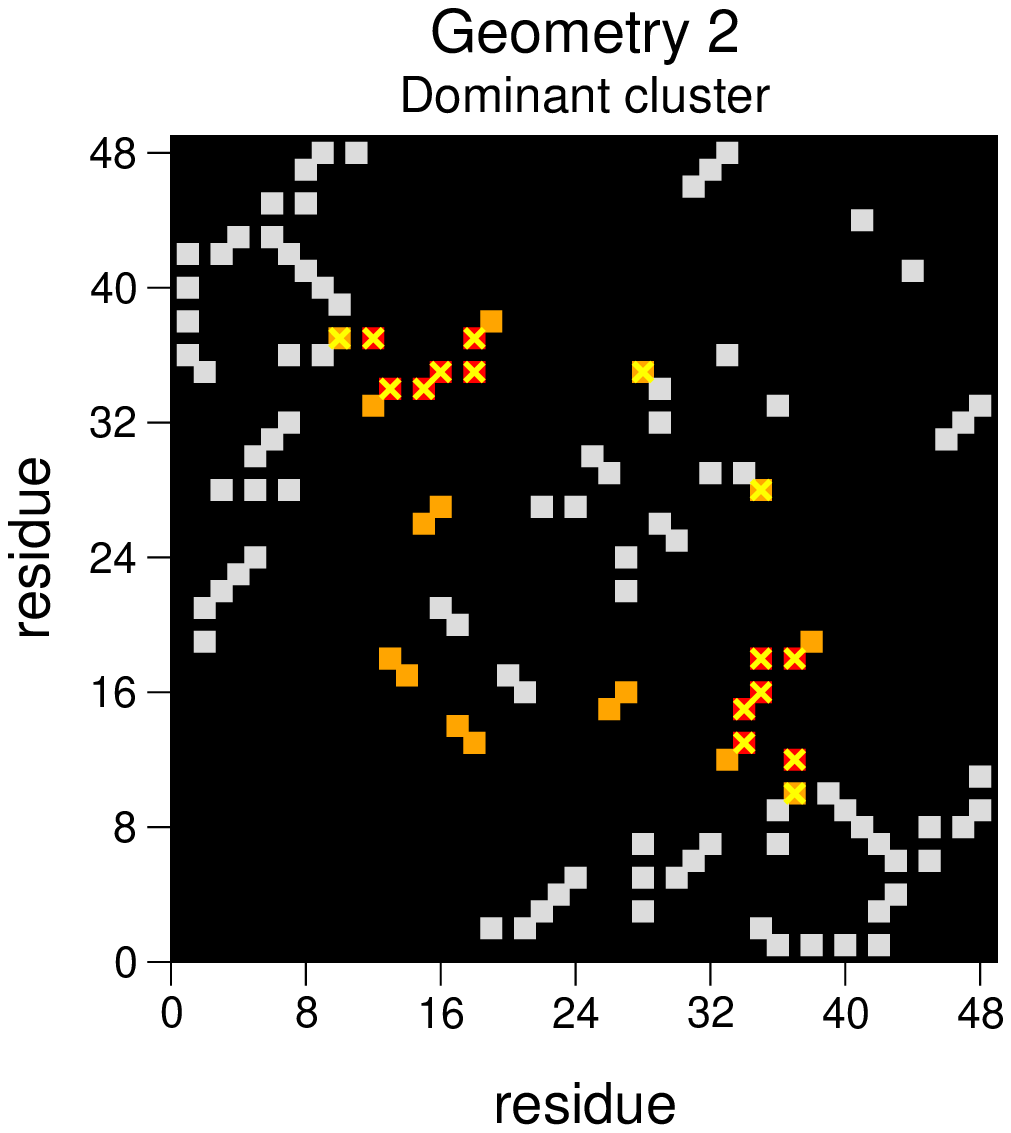}}}} \\
{\rotatebox{0}{\resizebox{7cm}{7cm}{\includegraphics{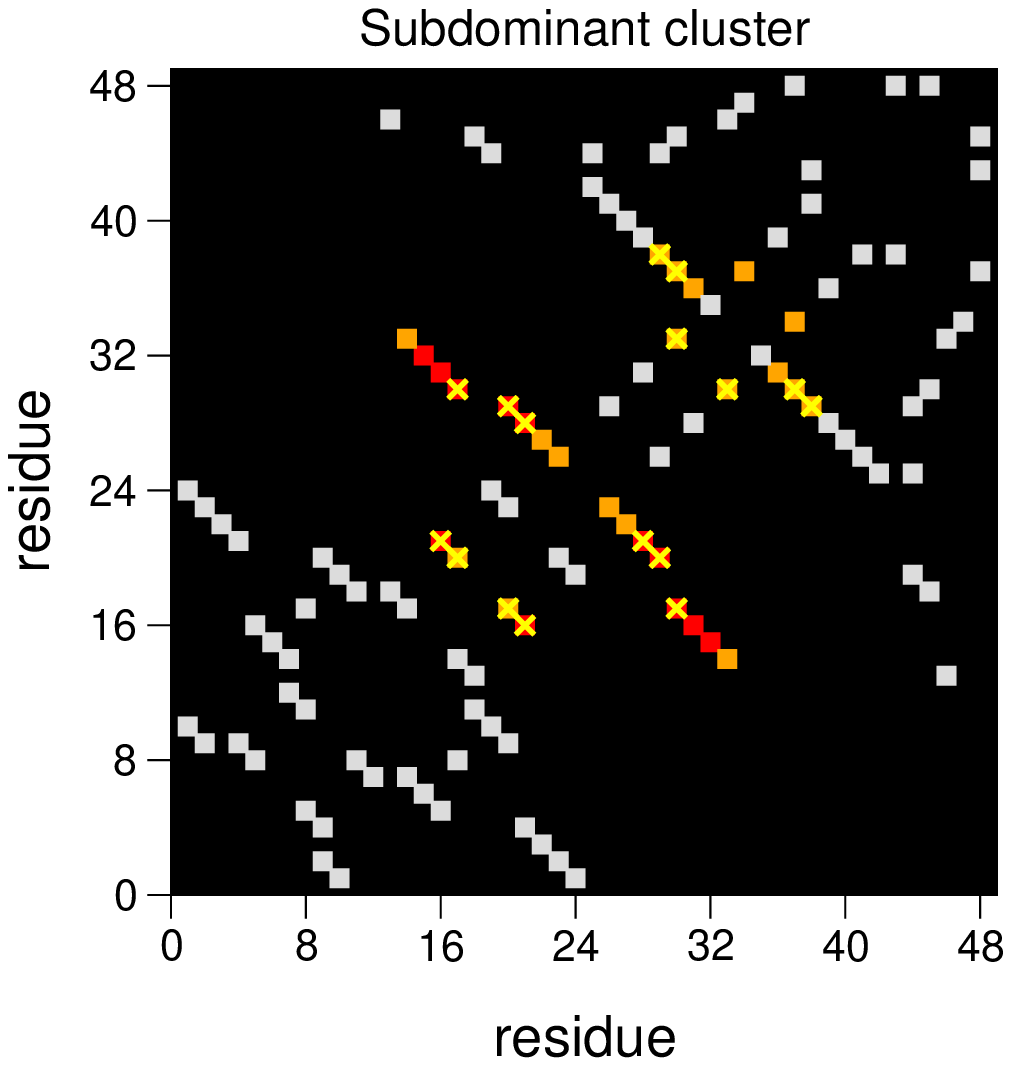}}}} 
{\rotatebox{0}{\resizebox{7cm}{7cm}{\includegraphics{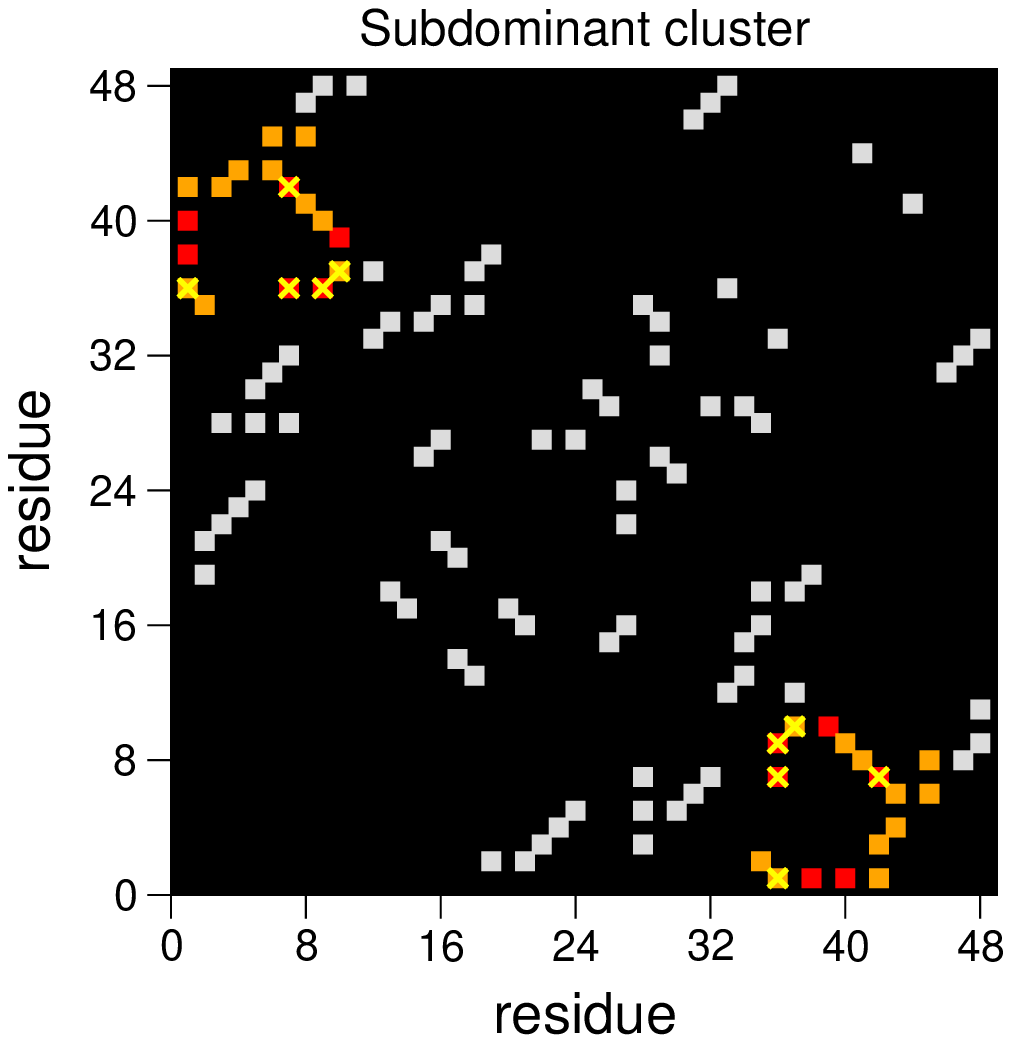}}}} 
\caption{Differential probability contact maps between pre- ($P_{fold}=0.05$) and TS ($P_{fold}=0.5$) conformations for geometry 1 (left) and geometry 2 (right). In the case
of geometry 1, the two structural classes identified within the dominant cluster, namely dominant clusters 1 and 2, are represented above and below the main diagonal of the corresponding contact map respectively. Native contacts that show a significant increase in contact probability are coloured  orange ($>50$\%), while those coloured red show the highest increase in contact probability (between $70\%$ and 95\% (85\%) for geometry 1 (2)). Native contacts in common with those identified as critical through $\phi$-value analysis are marked with a cross.}
\label{fig:no6}
\end{figure*}

\begin{figure*}
{\rotatebox{0}{\resizebox{5cm}{5cm}{\includegraphics{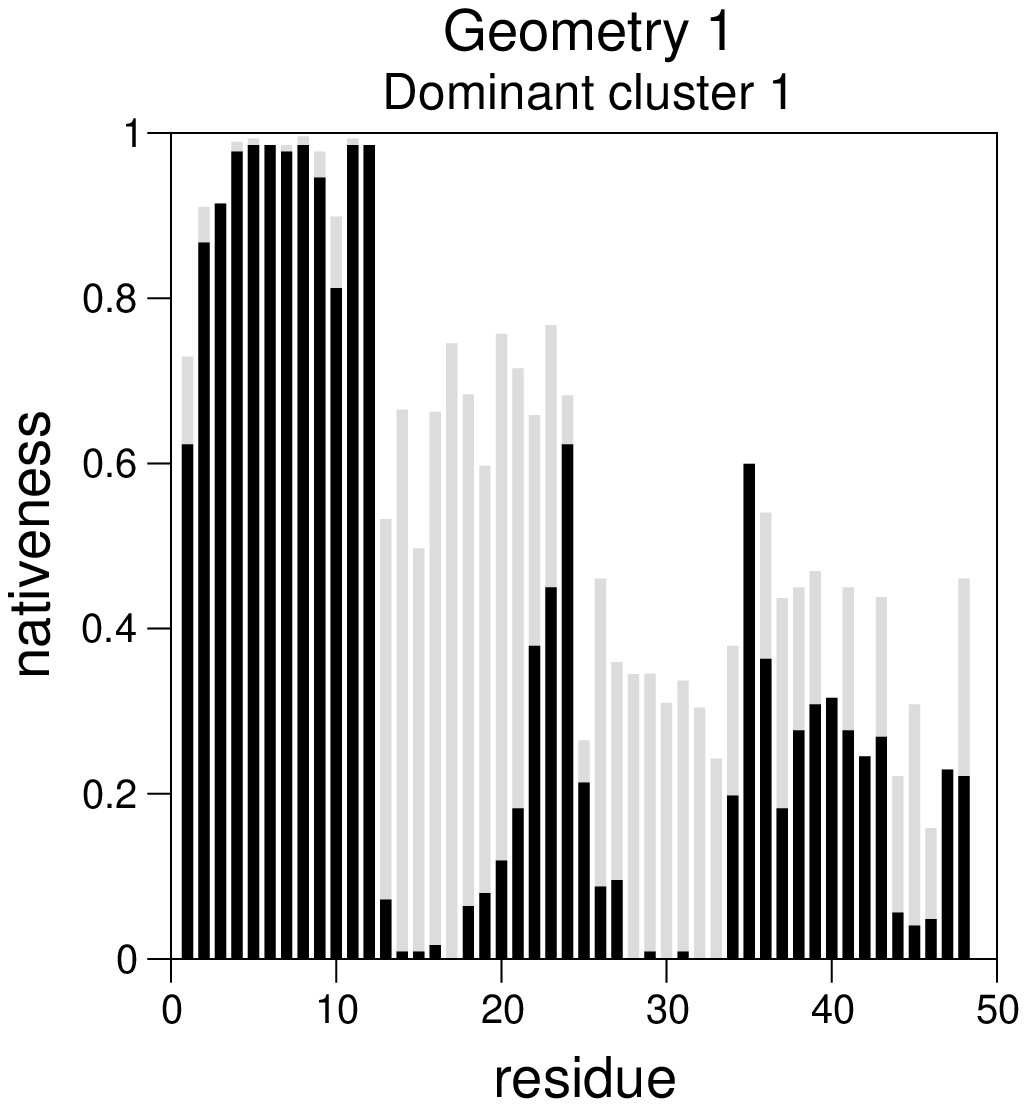}}}}
{\rotatebox{0}{\resizebox{5cm}{5cm}{\includegraphics{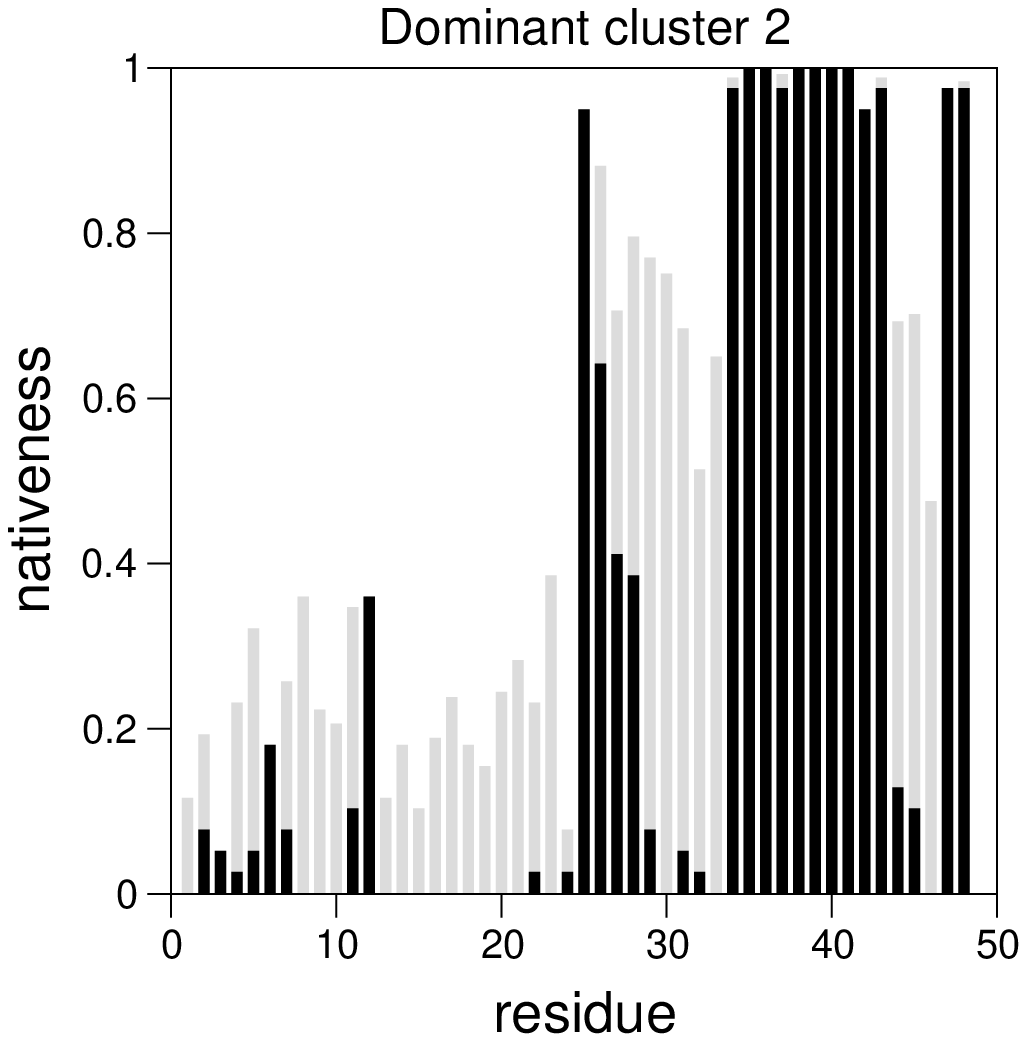}}}} 
{\rotatebox{0}{\resizebox{5cm}{5cm}{\includegraphics{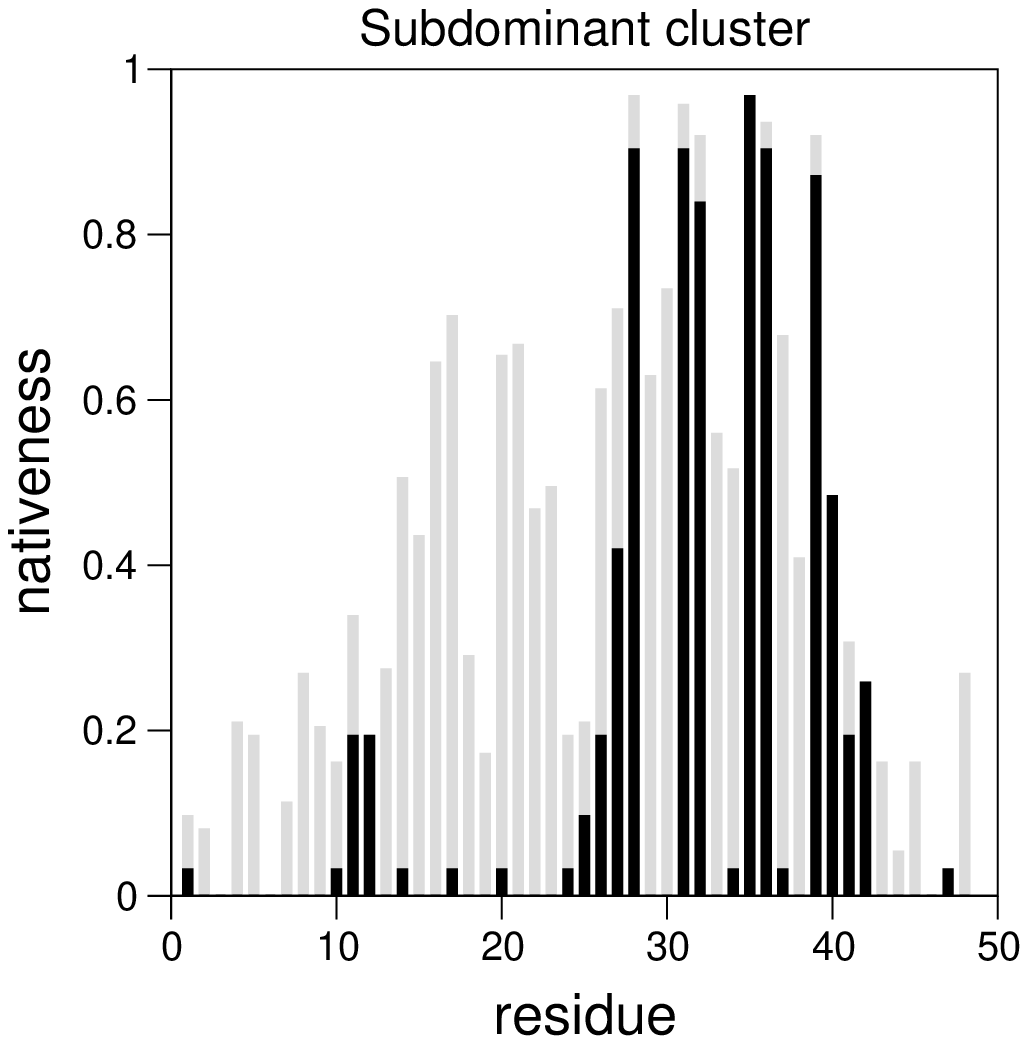}}}} \\
{\rotatebox{0}{\resizebox{5cm}{5cm}{\includegraphics{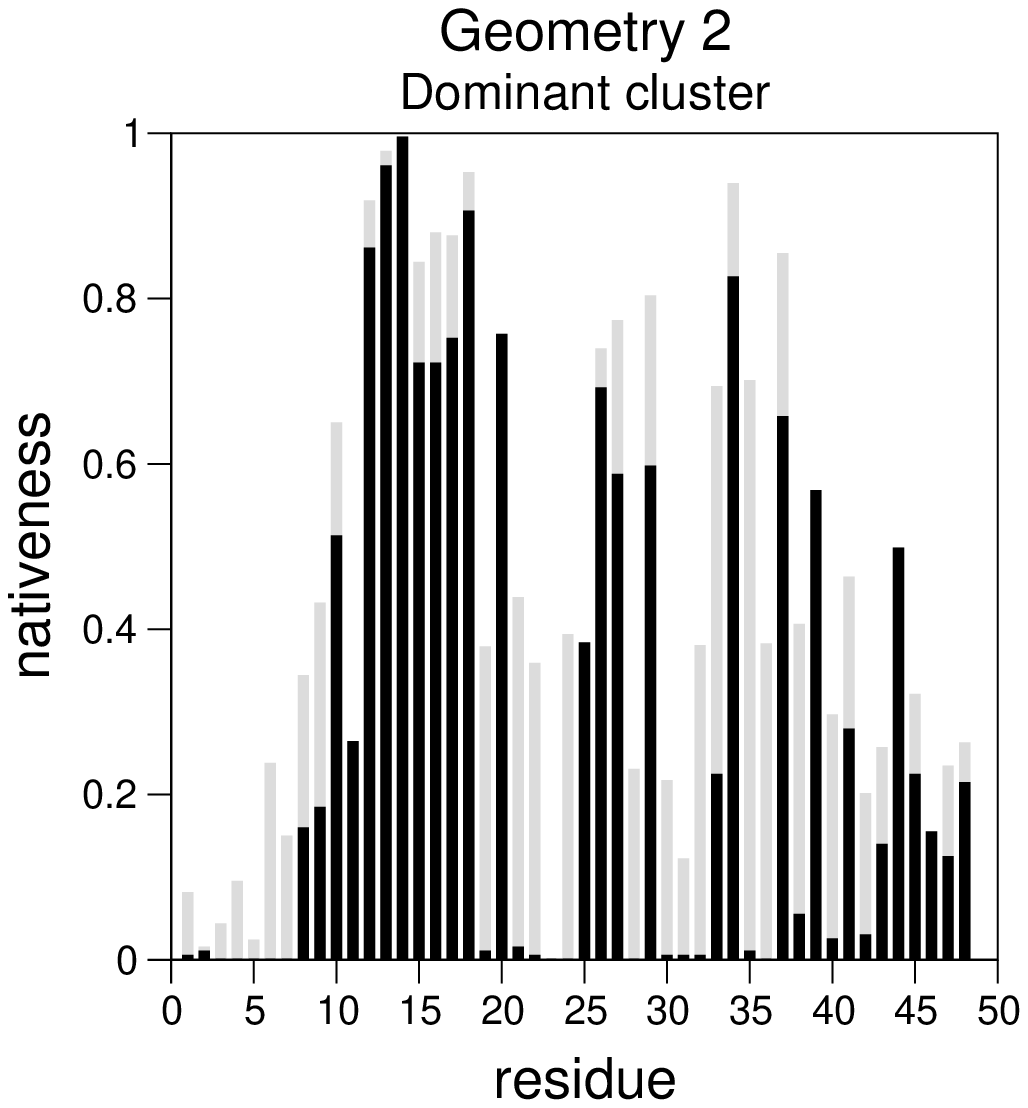}}}}
{\rotatebox{0}{\resizebox{5cm}{5cm}{\includegraphics{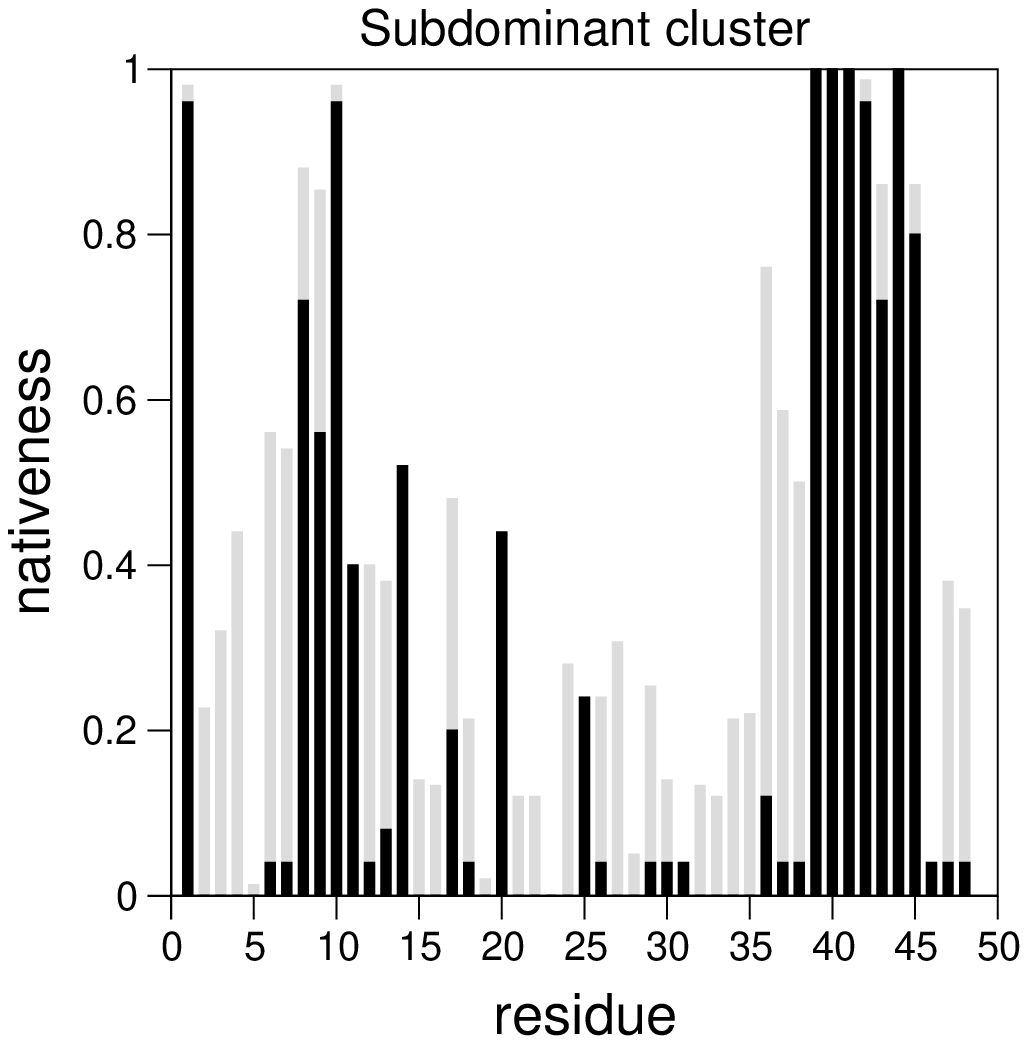}}}}

\caption{Probability that a residue is fully native (black) and mean fraction of native bonds established by each residue (grey) in the TS's dominant and subdominant clusters for geometry 1 (top row) and geometry 2 (bottom row).}
\label{fig:no7}
\end{figure*}

The bonds that exhibit the most dramatic changes between pre- and true TS conformations
are of key interest. Such bonds are comprised of the nucleus residues whose contacts both define and guarantee that the TS is reached~\cite{SHEA}. In order to determine which residues 
nucleate each identified TS structure we have determined the differential probability 
increase of each native bond between pre-TS conformations ($P_{fold}=0.05$) and the identified TS structures. Results reported in the differential probability maps (Fig.~\ref{fig:no6}) refer to the native bonds whose probability increase is higher than 50\%. Bonds that show a probability increase larger than 70\% (i.e., between 70\% and 95\% for geometry 1, and between 70\% and 85\% for geometry 2) are coloured red. A cross is used to mark the native bonds that are established by the residues identified as nucleating residues through $\phi$-value analysis. Interestingly, these are associated with the five native bonds that show the largest probability increases ($>$80\% for geometry 1, and $>$70\% for geometry 2) between pre- and TS structures. This finding strongly suggests that $\phi$-value analysis is able to pinpoint kinetically relevant residues independently of the change in the free energy of folding caused by mutation. For geometry 2, however, the $P_{fold}$ and $\phi$-value analysis give more consistent results than for geometry 1. Indeed, for geometry 2, there is a considerably larger overlap (75\%) than for geometry 1 (42\%) between the set of bonds identified as `key' bonds via $P_{fold}$ analysis (these are the bonds colored red in the differential probability maps) and the set of bonds established by the residues indentified as critical residues via the $\phi$-value analysis.

\subsubsection{Geometry 1}
In geometry 1 all the kinetically relevant residues 20 (via bond 20:29), 21 (via bonds 16:21 and 21:28), 29 (via bond 20:29) and 30 (via bond 17:30) nucleate the subdominant cluster. Residue 30 participates in the bond that shows the largest probability increase (93\%).
Dominant cluster 1 is nucleated by residues 20 (via bond 9:20) and 21 (via bonds 4:21 and 16:21) and, in this case, it is residue 21 the one that participates in the bond with the highest probability increase (90\%). Dominant cluster 2 is nucleated by residues 29 (via bond 29:44) and 30 (via bond 30:45). The very fact that the same residue can mediate the folding `reaction' through different pathways is suggestive of a unique transition state with microscopic heterogeneity.

\subsubsection{Geometry 2} 
In geometry 2 the structural class that is rich in long-range bonds (i.e., the subdominant cluster) is exclusively nucleated by residues 7 (via bonds 7:36 and 7:42) and 36 (via bonds 9:36 and 7:36). Residue 36 is associated with the two bonds whose probability increases the most ($>$74\%) between pre- and the TS conformations. On the other hand, residues 34 (via its bonds 13:34 and 15:34), 35 (via its bonds 18:35 and 16:35) and 37 (via its bonds 18:37 and 12:37 ) exclusively nucleate the dominant cluster. This suggests the existence of two structurally disjoint transition states.\par

\section{Critical residues and the structure of the transition state}

Here we investigate how the residues that are the determinant in the kinetics of folding are structured in the TS ensemble. In order to do so we measure the degree
of {\it nativeness} of every residue in each model protein by means of two different quantities.
One such quantity is the probability that a residue is fully native (i.e., has all its native bonds formed) in the TS, the other being the average fraction of native bonds formed by each residue in TS conformations. These two quantities are represented through the black and grey bars respectively in Figure~\ref{fig:no7} . \par
For geometry 1, all the four residues identified as being kinetically relevant have a vanishingly small probability of being fully native in all the identified TS structures. On average, however, residues 20 and 21 have 70\% of its native bonds formed in the dominant cluster 1, while residues 29 and 30 have about the same percent of bonds established in conformations pertaining to dominant cluster 2.  
In general, the dominant cluster is considerably more structured than the subdominant one. Indeed, residues 2 to 12, below the chain midpoint, have a probability larger than 81\% of being fully native in the dominant cluster 1, and residues 34 to 43, above the chain midpoint, have a very high probability $>$97\%  of being fully native in the dominant cluster 2. The subdominant cluster is more polarized: only residues 31, 32, 35 and 36 have all its native bonds established with a high probability ($>$84\%).\par
The TS of geometry 2 is considerably more polarized than that of geometry 1. Indeed, in this case  only residues 12, 13, 14 and 18 have a high probability $>$80\% of being fully native in the dominant cluster, and in the subdominant cluster it is residues 10, 39, 40, 41 and 44 that are fully native with a similarly high probability. Possibly due to topological constraints, one observes that residues 12 to 18, as well as residues 30 to 42, have on average more than 80\% of its native bonds formed. Except for residues 34 and 37, which have probabilities 75\% and 62\% of being fully native in the dominant cluster, the other kinetically relevant residues (namely residue 7, 35 and 36) have either a small or a vanishingly probability of having its all its native bonds formed in the TS.

\section{Conclusions}
 
The $\phi$-value analysis and other related methods~\cite{SOSNICK} are used as major tools to probe the structure of transition state (TS), and to identify the presence in this ensemble of the folding nucleus, i.e., the set of critical residues and associated native bonds that are the determinants of the  folding kinetics.\par 

Here we have employed a simulational proxy of the $\phi$-value analysis to identify the critical (i.e., nucleating) residues in two model proteins differing in native geometry. Results from extensive `mutagenesis' experiments, within the context of the lattice G\={o} model, revealed a set of residues whose mutation leads to a considerably large increase in the folding time. We found out that for the more complex protein geometry, which has predominantly non-local, long-range (LR) contacts, mutation of the critical residues has a much stronger impact on the folding time than for the geometry that is predominantly local. \par 
 
An advantage of computer simulations over {\it in vitro} with real proteins is the possibility to isolate and directly investigate the structure of TS conformations. The results of a thorough analysis, based on the reaction coordinate $P_{fold}$ and on the use of structural clustering, revealed a complex picture of the TS ensemble. Indeed, for both protein models the TS ensemble is heterogeneous, splitting up into subpopulations of structurally similar conformations. For the more complex geometry of the native structure the two identified populations are actually structurally disjoint, being associated with the existence of parallel folding pathways.\par 
For both geometries, the identification of the critical residues via the accurate   
$P_{fold}$ analysis agrees with the identification of the critical residues carried out with the  `$\phi$'-value analysis, which suggests that the latter can identify kinetically relevant residues in protein folding, independently of  the change in  freee energy of folding induced by mutation.  For the most complex geometry,  however, the $P_{fold}$ and $\phi$-value analysis give more consistent  results than for the more local geometry. This can be inferred from the overlap between the set of bonds identified as core critical bonds via the two considered methodologies, which is 30\% larger for the more complex 
geometry.\par 
The study of the TS structure reveals that the residues identified as critical through the `$\phi$'-value analysis are not necessarily fully native in neither of the identified TS ensemble subpopulations. Indeed, it is only for the more complex geometry that two of the five critical residues show a considerably high probability (up to 75\%) of having all its native bonds formed in the TS. Therefore, one concludes that in general the $\phi$-value correlates with the acceleration/deceleration of  folding induced  by mutation, rather than with the degree of nativeness of the TS~ \cite{DILL}, and that the `traditional' interpretation of $\phi$-values may provide a more accurate picture of the TS' structure only for more complex native geometries. 

Overall, our results suggest that native folds having predominantly non-local bonds are more suitable targets for $\phi$-value analysis than other protein geometries.\par

\section{\bf Acknowledgments}{P.F.N.F. thanks Funda\c c\~ao para a Ci\^encia e Tecnologia (FCT) for financial support through grants SFRH/BPD/21492/2005, POCI/QUI/58482/2004 and CRUP through grant B-7/05. R.D.M.T. thanks FCT for financial support through grant SFRH/BPD/27328/2006 and POCI/FIS/55592/2004. E.I.S. acknowledges support from the NIH grant GM52126.}

\clearpage

\section{\bf{Appendix}}

\begin{table*}
\caption{Mutations and resulting folding times observed for geometry 1. The folding time for the wild type sequence is $log_{10}(t)=5.64 \pm 0.04$. Also shown is the number of contacts disrupted (i.e.,  the number of interactions to each zero energy is ascribed) by each mutation.}
\begin{ruledtabular}
\begin{tabular}{c c c}
Mutation on bead(s)     &    \# contacts disrupted    &   $log_{10}(t)$    \\ \hline \hline
$33$ &    $3$ &     $5.87 \pm 0.04$   \\
$20$ &    $4$ &     $5.96 \pm 0.03$   \\
$21$ &    $3$ &     $5.97 \pm 0.04$   \\
$30$ &    $4$ &     $6.04 \pm 0.04$    \\
$29$ &    $4$ &     $6.11 \pm 0.04$    \\
\hline
$29, 30$ &    $8$ &  $6.64 \pm 0.04$   \\
$20, 33$ &    $7$ &  $6.65 \pm 0.04$   \\ 
$29, 20$ &    $7$ &  $6.87 \pm 0.04$   \\
$30, 21$ &    $7$ &  $6.92 \pm 0.04$   \\
$29, 21$ &    $7$ &  $6.97 \pm 0.04$   \\
$30, 20$ &    $8$ &  $7.27 \pm 0.05$   \\
\hline
$30, 20, 33$ &    $10$ &  $7.71 \pm 0.05$ \\ 
$30, 20, 29$ &    $11$ &  $7.77 \pm 0.05$  \\
$30, 21, 29$ &    $11$ &  $7.85 \pm 0.04$  \\
$30, 20, 21$ &    $11$ &  $8.05 \pm 0.05$ \\ 
\end{tabular}
\label{table_mutG1}
\end{ruledtabular}
\end{table*}

\begin{table*}
\caption{Mutations and resulting folding times observed for geometry 2. The folding time for the  wild type sequence is $log_{10}(t)=6.29 \pm 0.05$. Also shown is the number of contacts disrupted (i.e., the number of interactions to which zero energy is ascribed) by each mutation. The folding time of the triple-point mutations marked with an * is estimated. Indeed, for the triple mutants 7:35:36 and 7:35:37  only 71\% and 56\% of the MC runs respectively reached the native state in the allowed number of MC steps.}
\begin{ruledtabular}
\begin{tabular}{c c c}
Mutation on bead(s)     &    \# contacts disrupted     &   $log_{10}(t)$    \\ \hline \hline

$8$ &     $3$ &     $6.62 \pm 0.04$   \\    
$28$ &    $4$ &     $6.63 \pm 0.04$   \\    
$6$ &     $3$ &     $6.68 \pm 0.06$   \\    
$10$ &    $2$ &     $6.72 \pm 0.05$   \\
$19$ &    $4$ &     $6.76 \pm 0.05$   \\
$21$ &    $2$ &     $6.76 \pm 0.05$   \\
$34$ &    $3$ &     $6.78 \pm 0.05$   \\
$9$ &     $3$ &     $6.83 \pm 0.05$   \\   
$34$ &    $1$ &     $6.78 \pm 0.05$   \\
$35$ &    $2$ &     $6.82 \pm 0.04$   \\
$7$ &     $4$ &     $6.84 \pm 0.05$   \\
$37$ &    $3$ &     $6.84 \pm 0.05$   \\
$36$ &    $4$ &     $7.10 \pm 0.05$    \\
\hline
$37, 34$ &    $6$ &  $7.24 \pm 0.05$   \\
$36, 9$ &     $7$ &  $7.31 \pm 0.04$   \\
$9, 37$ &     $6$ &  $7.33 \pm 0.04$   \\
$9, 34$ &     $6$ &  $7.42 \pm 0.05$   \\
$35, 37$ &    $7$ &  $7.46 \pm 0.04$   \\
$36, 34$ &    $7$ &  $7.51 \pm 0.05$   \\
$36, 37$ &    $7$ &  $7.52 \pm 0.04$   \\
$36, 7$ &     $7$ &  $7.56 \pm 0.05$   \\
$7, 35$ &     $7$ &  $7.60 \pm 0.04$   \\
$7, 34$ &     $7$ &  $7.73 \pm 0.05$   \\
$7, 37$ &     $7$ &  $7.76 \pm 0.04$   \\
$35, 36$ &    $7$ &  $7.80 \pm 0.04$   \\
\hline
$36,37,34$ &    $10$ &  $8.00 \pm 0.04$ \\
$7,34,36$ &     $10$ & $8.45 \pm 0.04$ \\
$7,37,34$ &     $10$ & $8.58 \pm 0.04$ \\
$7,35,36^*$ &     $11$ &  $9.46 \pm 0.06$ \\ 
$7,35,37^*$ &     $10$ & $9.61 \pm 0.05$ \\ 
\end{tabular}
\label{table_mutG2}
\end{ruledtabular}
\end{table*}

\begin{table*}
\caption{Structural clusters identified for geometry 1 along the reaction coordinate $P_{fold}$.
$<Q>$ is the average fraction of native contacts formed in the starting ensemble, D stands for dominant cluster, SD for subdominant cluster, and T represents the third cluster that emerges from
$P_{fold}=0.6$ onwards. The (averaged mean) time to fold starting from the conformations in each cluster is shown as a fraction of the folding time starting from a random coil conformer.}
\begin{ruledtabular}
\begin{tabular}{c c c c c}
$P_{fold}$ &   \# conformations    &      $<Q>$     &       cluster (\#conformations) & time to fold (\%MFPT) \\ \hline \hline
 0.2   &       $598$           &    $0.45$    &             D1:108; D2:28      &           (41\%, 35\%)\\
 0.3   &       $498$           &    $0.48$    &             D:240; SD:65 &           (33\%, 28\%)\\
 0.4   &       $450$           &    $0.49$    &             D1:133; D2:48; SD:21&        (32\%,  28\%,     17\%)\\
 0.5   &       $452$           &    $0.50$    &             D1:127; D2:39; SD:31&         (33\%, 25\%,  17\%)\\ 
 0.6   &       $427$           &    $0.51$    &             D:111; T:57; SD:41&          (21\%, 41\%, 17\%) \\
 0.7   &       $541$           &    $0.54$    &             D:215; T:67; SD:73&          (28\%, 22\%, 17\%)            \\  
 0.8  &       $1170$          &     $0.59$    &             D:674; T:170; SD:71 &        (17\%, 30\%, 16\%)\\         
\end{tabular}
\label{clusters1}
\end{ruledtabular}
\end{table*}

\begin{table*}
\caption{Structural clusters identified for geometry 2 along the reaction coordinate $P_{fold}$.
$<Q>$ isthe average fraction of native contacts formed in the starting ensemble, D stands for dominant cluster, SD for subdominant cluster, and Trp represents a trapped state.}
\begin{ruledtabular}
\begin{tabular}{c c c c c}
$P_{fold}$ &   \# conformations    &      $<Q>$     &       cluster:\#conformations & time to fold (\%MFPT)\\ \hline \hline
 0.2   &       $482$           &    $0.33$    &           D:215; SD:27             &  (22\%, 16\%) \\
 0.3   &       $443$           &    $0.35$    &           D:207; SD:38         &    (50\%, 14\%)\\ 
 0.4   &       $406$           &    $0.39$    &           D:219; SD:46         &   (58\%, 8.8\%)\\
 0.5   &       $401$           &    $0.41$    &           D:228; SD:55         &    (31\%, 7.7\%)\\
 0.6   &       $337$           &    $0.40$    &           D:148; SD:24         &   (19\%, 7.7\%)\\ 
 0.7   &       $338$           &    $0.42$    &           D:119; SD:29         &    (16\%, 7.4\%)\\ 
 0.8   &       $449$           &    $0.46$    &           D:117; SD:76; Trp:27 &   (11\%, 6.2\%, 49\%)\\
\end{tabular}
\label{clusters2}
\end{ruledtabular}
\end{table*}

\begin{table*}
\caption{Number of conformations in the different clusters (dominant, subdominant and third cluster) at $P_{fold}=0.8$ as a function of the starting conformations for geometry 1.}
\begin{ruledtabular}
\begin{tabular}{c c c c}
Starting conformations&   D at $P_{fold}=0.80$   &  T at $P_{fold}=0.95$ & SD at $P_{fold}=0.80$    \\ \hline \hline
 Unfolded State   &       674/915 = 74\%           &      170/915 = 19\%   &      71/915 = 7.8\%  \\
 D1 at $P_{fold}=0.5$   &        239/255 = 93\%       &   16/255 = 6.6\%  & 0\%  \\
 D2 at $P_{fold}=0.5$   &        13/78 = 17\%           &      65/78 = 83\%  & 0\%  \\
 SD at $P_{fold}=0.5$   &       100/133 = 75\%           &      14/133 = 11\% & 19/133 = 19\%   \\
\end{tabular}
\label{runclusters1}
\end{ruledtabular}
\end{table*}

\begin{table*}
\caption{Number of conformations in the different clusters (dominant, subdominant and trapped state) at $P_{fold}=0.8$ as a function of the starting conformations for geometry 2.}
\begin{ruledtabular}
\begin{tabular}{c c c c}
Starting conformations&   D at $P_{fold}=0.8$  &    SD at $P_{fold}=0.8$   &    Trp at $P_{fold}=0.8$   \\ \hline \hline
 Unfolded State   &       117/220 = 53\%           &      76/220 = 35\%   &   27/220=12\%\\
 D at $P_{fold}=0.5$   &       44/51 = 90\%        &  0\%     & 5/51 = 10\%\\
 SD at $P_{fold}=0.5$   &       0\%           &      100/100 = 100\% & 0\%  \\
\end{tabular}
\label{runclusters2}
\end{ruledtabular}
\end{table*}

\clearpage

\clearpage

\end{document}